\documentclass[numberedappendix,appendixfloats]{emulateapj}
\usepackage{url}
\usepackage{color}
\shorttitle{Stellar Evolution on the TP-AGB}
\shortauthors{Kalirai, Marigo, \& Tremblay}

\begin{document}

\title{The Core Mass Growth and Stellar Lifetime 
of Thermally \\ Pulsing Asymptotic Giant Branch Stars}

\author{
Jason S.\ Kalirai\altaffilmark{1,2}, 
Paola Marigo\altaffilmark{3}, \&
Pier-Emmanuel Tremblay\altaffilmark{1,4,5}}
\altaffiltext{1}{Space Telescope Science Institute, 3700 San Martin Drive, Baltimore, 
MD, 21218; jkalirai@stsci.edu}
\altaffiltext{2}{Center for Astrophysical Sciences, Johns Hopkins University, Baltimore, MD, 21218}
\altaffiltext{3}{Department of Physics and Astronomy, University of Padova, Vicolo dell'Osservatorio 3, 35122 Padova, 
Italy; paola.marigo@unipd.it}
\altaffiltext{4}{Zentrum f\"ur Astronomie der Universit\"at Heidelberg, Landessternwarte, 
K\"onigstuhl 12, D-69117 Heidelberg, Germany; ptremblay@lsw.uni-heidelberg.de}
\altaffiltext{5}{Hubble Fellow}


\begin{abstract}

\noindent

We establish new constraints on the intermediate-mass range of the initial-final mass 
relation by studying white dwarfs in four young star clusters, and apply the results to 
study the evolution of stars on the thermally pulsing 
asymptotic giant branch (TP-AGB).  We show that the stellar core mass on the AGB 
grows rapidly from 10\% to 30\% for stars with $M_{\rm initial}$ = 1.6 to 2.0~$M_\odot$.  At 
larger masses, the core-mass growth decreases steadily to $\sim$10\% at $M_{\rm initial}$ = 
3.4~$M_\odot$.  These observations are in excellent agreement with predictions from the 
latest TP-AGB evolutionary models in \cite{marigo13}.  We also compare to models 
with varying efficiencies of the third dredge-up and mass loss, and demonstrate 
that the process governing the growth of the core is largely the stellar wind, 
while the third dredge-up plays a secondary, but non-negligible role. 
Based on the new white dwarf measurements, we perform an exploratory calibration
of the most popular mass-loss prescriptions in the literature.  Finally, we estimate the 
lifetime and the integrated luminosity of stars on the TP-AGB to peak at $t$ $\sim$ 3~Myr and 
$E$ = 1.2 $\times$ 10$^{10}$~$L_\odot$~yr for $M_{\rm initial}$ $\sim$ 2~$M_\odot$ ($t$ $\sim$ 2~Myr 
for luminosities brighter than the RGB tip at $\log(L/L_{\odot})$ $>$ 3.4), decreasing to $t$ = 
0.4~Myr and $E$ = 6.1 $\times$ 10$^{9}$~$L_\odot$~yr for stars with $M_{\rm initial}$ $\sim$ 
3.5~$M_\odot$.  The implications of these results are discussed with respect to general 
population synthesis studies that require correct modeling of the TP-AGB phase of stellar 
evolution.

%
%

\end{abstract}

\keywords{open clusters and associations: individual (Hyades and Praesepe) - 
stars: evolution, AGB and post-AGB - techniques: photometric, spectroscopic - white dwarfs}


\section{Introduction} \label{introduction}

The life cycles of most stars are dominated by quiescent, long-lived phases  
such as the hydrogen-burning main sequence and the white dwarf cooling sequence.  
For low- and intermediate-mass stars with initial masses in the range  
1~$M_\odot$ $\gtrsim$ $M_{\rm initial}$ $\gtrsim$ 6 -- 8~$M_\odot$, these two extremes 
are connected by  the thermally pulsing asymptotic giant branch (TP-AGB) evolutionary phase, 
during which stars experience quasi periodic thermal instabilities of the He-burning shell 
(thermal pulses) and rapidly lose a large fraction of their mass \citep{herwig05}.

An understanding of the TP-AGB phase has many important 
applications in astronomy.  Of particular interest is the prospect of directly 
measuring the growth of the stellar core on the AGB. The growth 
is set by the lifetime of the TP-AGB, which itself depends on the timescale over which the 
stellar envelope is lost through mass loss processes \citep{marigogirardi_01}.
At the same time, the effective increase of the core may be limited by the third dredge-up, 
which causes a sudden reduction of its mass each time it takes place \citep{herwig04}.
This growth of the core mass and the TP-AGB lifetime as a function of the initial stellar mass 
(hence age) are powerful inputs to theoretical models aimed at evaluating the integrated 
luminosity contribution of AGB stars,  since these luminosities play a central role in the 
construction of population synthesis models that are used to interpret galaxy 
evolution (e.g., Bruzual \& Charlot 2003; Maraston et~al.\ 2006; 
Conroy 2009: Conroy \& Gunn 2010; Zibetti et~al.\ 2013).


\begin{table*}
\begin{center}
\caption{Hyades and Praesepe Cluster White Dwarfs}
\begin{tabular}{lcccccc}
\hline
\hline
\multicolumn{1}{c}{Cluster} & \multicolumn{1}{c}{ID} & 
\multicolumn{1}{c}{$T_{\rm eff}$ (K)} & \multicolumn{1}{c}{log~$g$} & \multicolumn{1}{c}{$M_{\rm final}$ ($M_\odot$)} &
\multicolumn{1}{c}{log($t_{\rm cool})$ (yr)} & \multicolumn{1}{c}{$M_{\rm initial}$ ($M_\odot$)} \\
\hline
NGC6819 &  NGC6819\_6 & 21,900 $\pm$ 300 & 7.89 $\pm$ 0.04 & 0.56 $\pm$ 0.02 & 7.56 $\pm$ 0.04   & $1.60^{+0.06}_{-0.05}$ \\ 
NGC6819 &  NGC6819\_7 & 16,600 $\pm$ 200 & 7.97 $\pm$ 0.04 & 0.59 $\pm$ 0.02 & 8.14 $\pm$ 0.04   & $1.62^{+0.07}_{-0.05}$ \\ 
NGC7789 &  NGC7789\_5 & 31,600 $\pm$ 200 & 7.98 $\pm$ 0.05 & 0.64 $\pm$ 0.03 & 6.95 $\pm$ 0.05   & $2.02^{+0.07}_{-0.14}$ \\     
NGC7789 &  NGC7789\_8 & 25,000 $\pm$ 400 & 8.06 $\pm$ 0.07 & 0.66 $\pm$ 0.04 & 7.46 $\pm$ 0.07   & $2.02^{+0.09}_{-0.11}$ \\ 
Hyades & WD0352+096 &  14,670 $\pm$ 380 & 8.30 $\pm$ 0.05 & 0.80 $\pm$ 0.03 & 8.53 $\pm$ 0.05    & $3.59^{+0.21}_{-0.15}$ \\ 
Hyades & WD0406+169 &  15,810 $\pm$ 290 & 8.38 $\pm$ 0.05 & 0.85 $\pm$ 0.03 & 8.50 $\pm$ 0.04    & $3.49^{+0.13}_{-0.10}$ \\
Hyades & WD0421+162 &  20,010 $\pm$ 320 & 8.13 $\pm$ 0.05 & 0.70 $\pm$ 0.03 & 7.97 $\pm$ 0.06    & $2.90^{+0.02}_{-0.02}$ \\
Hyades & WD0425+168 &  25,130 $\pm$ 380 & 8.12 $\pm$ 0.05 & 0.71 $\pm$ 0.03 & 7.49 $\pm$ 0.08    & $2.79^{+0.01}_{-0.01}$ \\
Hyades & WD0431+126 &  21,890 $\pm$ 350 & 8.11 $\pm$ 0.05 & 0.69 $\pm$ 0.03 & 7.78 $\pm$ 0.07    & $2.84^{+0.02}_{-0.02}$ \\
Hyades & WD0437+138 &  15,120 $\pm$ 360 & 8.25 $\pm$ 0.09 & 0.74 $\pm$ 0.06 & 8.47 $\pm$ 0.07    & $3.41^{+0.21}_{-0.15}$ \\
Hyades & WD0438+108 &  27,540 $\pm$ 400 & 8.15 $\pm$ 0.05 & 0.73 $\pm$ 0.03 & 7.30 $\pm$ 0.09    & $2.78^{+0.01}_{-0.01}$ \\
Hyades & WD0348+339 &  14,820 $\pm$ 350 & 8.31 $\pm$ 0.05 & 0.80 $\pm$ 0.03 & 8.52 $\pm$ 0.05    & $3.55^{+0.19}_{-0.14}$ \\
Hyades & HS0400+1451 &  14,620 $\pm$  60 & 8.25 $\pm$ 0.01 & 0.76 $\pm$ 0.01 & 8.50 $\pm$ 0.01    & $3.49^{+0.03}_{-0.03}$ \\
Hyades & WD0625+415 &  17,610 $\pm$ 280 & 8.07 $\pm$ 0.05 & 0.66 $\pm$ 0.03 & 8.12 $\pm$ 0.05    & $2.97^{+0.03}_{-0.03}$ \\
Hyades & WD0637+477 &  14,650 $\pm$ 590 & 8.30 $\pm$ 0.06 & 0.80 $\pm$ 0.04 & 8.53 $\pm$ 0.06    & $3.59^{+0.26}_{-0.18}$ \\
Praesepe & WD0833+194 & 15,252 $\pm$ 41 & 8.28 $\pm$ 0.01 & 0.79 $\pm$ 0.04 & 8.47 $\pm$ 0.05    & $3.41^{+0.16}_{-0.09}$ \\
Praesepe & WD0836+199 & 14,971 $\pm$ 60 & 8.33 $\pm$ 0.01 & 0.82 $\pm$ 0.04 & 8.53 $\pm$ 0.05    & $3.59^{+0.18}_{-0.13}$ \\
Praesepe & WD0837+185 & 15,476 $\pm$ 60 & 8.41 $\pm$ 0.01 & 0.87 $\pm$ 0.04 & 8.55 $\pm$ 0.05    & $3.66^{+0.21}_{-0.16}$ \\
Praesepe & WD0837+199 & 17,640 $\pm$ 38 & 8.30 $\pm$ 0.01 & 0.80 $\pm$ 0.04 & 8.30 $\pm$ 0.05    & $3.13^{+0.06}_{-0.05}$ \\
Praesepe & WD0840+190 & 15,335 $\pm$ 68 & 8.48 $\pm$ 0.01 & 0.91 $\pm$ 0.05 & 8.61 $\pm$ 0.05    & $3.97^{+0.40}_{-0.24}$ \\
Praesepe & WD0840+200 & 15,383 $\pm$ 42 & 8.28 $\pm$ 0.01 & 0.79 $\pm$ 0.04 & 8.46 $\pm$ 0.05    & $3.39^{+0.12}_{-0.09}$ \\
Praesepe & WD0843+184 & 15,418 $\pm$ 50 & 8.44 $\pm$ 0.01 & 0.89 $\pm$ 0.05 & 8.57 $\pm$ 0.05    & $3.77^{+0.27}_{-0.18}$ \\
\hline
\end{tabular}
\label{table1}
\end{center}
\end{table*}


%
%


On the other hand it is a matter of fact that,
in spite of the remarkable progress attained in fields of TP-AGB stellar 
evolution in the last decades \citep[see][for a review]{herwig05},  predictions of this phase
are still affected by a sizable degree of uncertainty.
This should be mostly ascribed to  the high complexity of the physics involved, and the fact  
we still have to cope with  ill-defined theories of stellar mixing and convection, 
as well as insufficient understanding of mass loss mechanisms.
We still lack an accurate knowledge of  how the third dredge-up episodes 
vary with thermal pulses, and of what is the dependence of their 
efficiency on  stellar mass  and metallicity.  Likewise, substantial
effort is needed to gain insight into the driving mechanism  and strengths of 
stellar winds on the AGB (e.g., Habing~1996; Weidemann~2000; 
Willson~2000; Gustaffsson \& H\"ofner 2004).  The relation between mass loss and 
other stellar parameters such as metallicity and dust-to-gas ratio is also not well understood.  
Similarly, direct observational constraints are difficult to establish given the dust 
enshrouded nature of AGB stars and their short evolutionary lifetimes.

The relation between the initial and final (i.e., white dwarf) masses of stars 
represents a new tool to bear on studies of AGB evolution (Bird \& Pinsonneault 
2011), since the end product of AGB stars is the white dwarf cooling sequence 
(e.g., Weidemann~2000; Girardi et~al.\ 2010).  The relation has now been 
well-measured by spectroscopically studying white dwarfs that are members of star 
clusters with well defined characteristics.  The current constraints from $M$ = $\sim$1 -- 
7~$M_\odot$ shows a rise in the remnant mass that is proportional to the initial 
mass (e.g., see Kalirai et~al.\ 2007; 2008; 2009 and references therein).  At the intermediate 
masses that are characteristic of AGB stars, the relation exhibits a large scatter and 
this leads to difficulty in ascertaining the influence of AGB evolution.  This scatter is likely 
caused by the heterogeneous nature of previous studies.  The white dwarf 
spectra have been collected with different instrumentation and suffer from many selection 
effects and biases.  For example, there is likely contamination in the sample from 
field stars, low signal-to-noise ratio measurements, fits to Balmer lines using outdated 
spectroscopic models, incorrect metallicity assumptions, and inaccurate turnoff ages 
inferred from different theoretical isochrones (leading to systematic errors in the 
initial masses). 

\cite{bird11} recently investigated the initial-final mass relation and employed a 
fuel consumption argument to set a lower bound on the fraction of light emitted 
during the TP-AGB phase.  Their results, based on combining several studies of the 
initial-final mass relation, suggest that the growth of the stellar core exhibits a 
plateau of $\sim$20\% at $M_{\rm initial}$ $\sim$ 3~$M_\odot$, decreasing to $\sim$10\% 
at $M_{\rm initial}$ $>$ 4~$M_\odot$.  In the present study, we build on the initial work by 
\cite{bird11} by taking advantage of new observational and theoretical work.  First, 
we minimize systematic errors by limiting our study to a small set of star clusters that all 
have moderately super-Solar metallicity, two of which also have identical ages.\footnote{The 
nine star clusters in \cite{bird11} spanned a metallicity range of greater than a factor of 
2.}  Second, we take advantage of newly discovered white dwarfs in both the Hyades and 
Praesepe star clusters to increase the significance of the measurement over the critical 
mass range that corresponds to expected AGB evolution.  Finally, we largely 
eliminate systematic errors in the derivation of remnant masses by re-calculating 
all measurements with a common set of white dwarf spectral models that incorporate the 
latest physics of the Stark broadening.  The result of this work is a robust measurement 
of the core-mass growth at $M_{\rm initial}$ = 1.6 to 3.8~$M_\odot$. 

We describe the observational data set in \S\,\ref{sec:obs} and the calculation of initial 
and final masses for each star in \S\,\ref{sec:masses}.  These results provide new constraints 
on the absolute core-mass growth of the AGB (\S\,\ref{sec:coremassgrowth}), the processes 
governing core-mass growth including the significance of the third dredge up (\S\,\ref{sec_3dup}) 
and mass loss (\S\,\ref{sec_mloss}), and the lifetime and energy output of these 
stars (\S\,\ref{sec:lum}).  All of the results are discussed with respect to the important role 
that the TP-AGB phase of stellar evolution plays in establishing the fraction of red light 
emitted in population synthesis models.

\begin{figure*}[ht]
\begin{center}
\leavevmode 
\includegraphics[width=12.0cm]{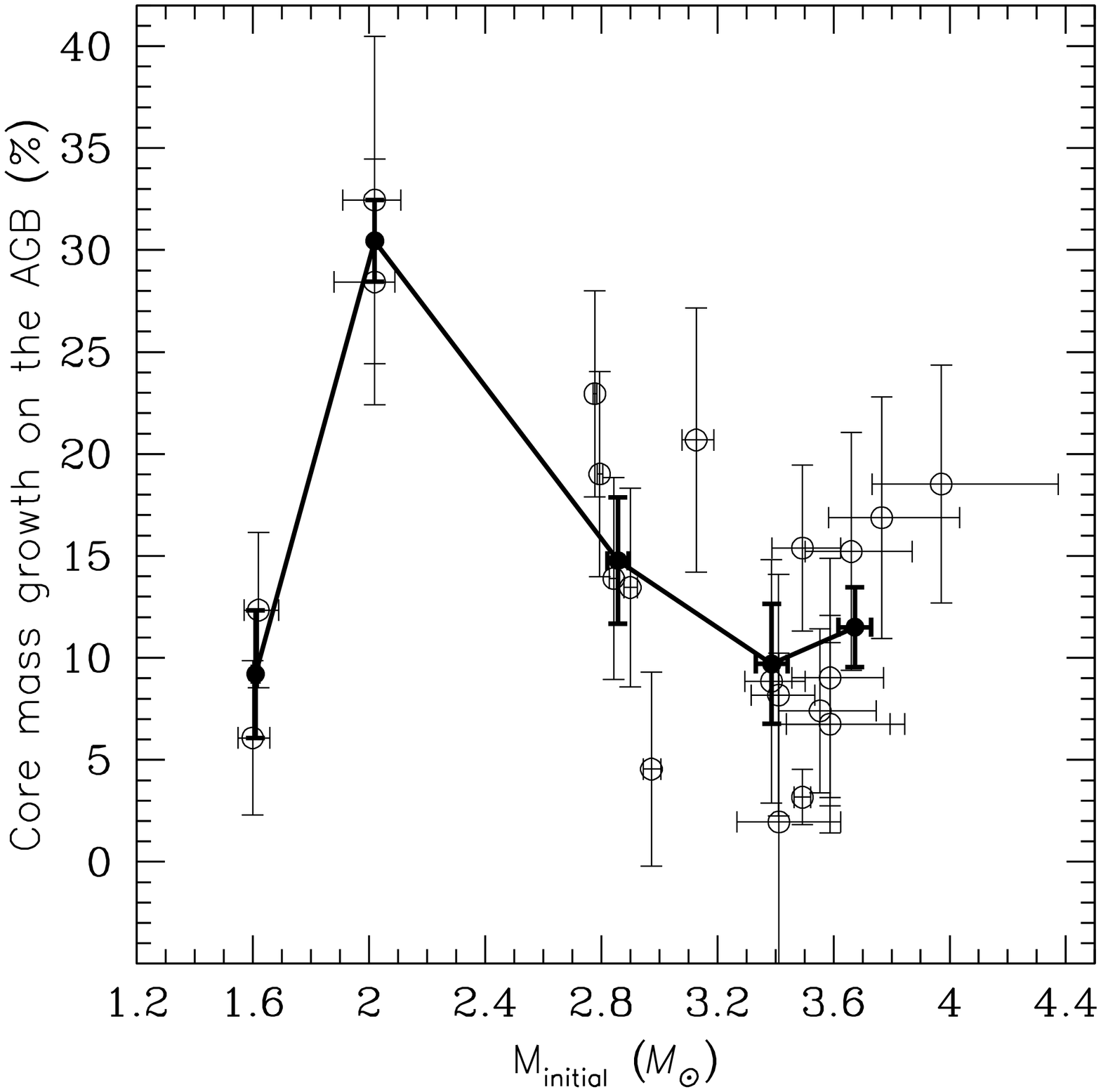}
\end{center}
\vspace{-0.4cm}
\caption{The growth of the stellar core on the TP-AGB ($\Delta M_{\rm growth}$ = $M_{\rm final}$ $-$ $M_{\rm{ c, 1tp}}$), 
measured by comparing the masses of bright white dwarfs in four star clusters 
($M_{\rm final}$ -- see \S\,\ref{sec:masses}) to the core mass at the first thermal pulse from the 
new \cite{bressan12} stellar models ($M_{\rm{ c, 1tp}}$ -- see \S,\ref{sec:coremassgrowth}).  The data 
points with error bars illustrate individual measurements in the four star clusters, and the solid black 
line shows average values (and errors in the averages) across five $M_{\rm initial}$ bins as described in 
Section~\ref{sec:coremassgrowth}.  The maximum growth of the stellar core of AGB stars occurs for stars 
with $M_{\rm initial}$ $\sim$ 2.0~$M_\odot$. \label{fig:growth}}
\end{figure*}


\section{New White Dwarfs in the Hyades and Praesepe Star Clusters} \label{sec:obs}

The Hyades and Praesepe open star clusters share incredible similarities.  Both clusters 
have ages of $\sim$600 -- 650~Myr and metallicities slightly higher than Solar, $Z_{\rm initial} \sim$ 0.02
(Gratton~2000; An et~al.\ 2008), and can be studied in exquisite detail given their proximity 
($d$ = 46.3~pc for Hyades -- Perryman et~al.\ 1998; $d$ = 184.5~pc for Praesepe -- An et~al.\ 2008).  
The present main-sequence turnoff mass in these clusters is $\sim$3~$M_\odot$.  

Recent observations of both the Hyades and Praesepe clusters have revealed new members of the 
remnant white dwarf population.  For the Hyades, \cite{schilbach12} constructed a multi-step 
process to identify 27 white dwarf {\it candidates}, including all 10 of the previously 
known members (van Altena 1969; Reid~1992; Weidemann et~al.\ 1992; von Hippel~1998).  
Their methods combine tangential motions from proper motion measurements, photometric 
comparisons with the white dwarf locus in the color-magnitude diagram, and radial 
velocities for some stars.  \cite{tremblay12} further scrutinized the membership of 
these candidates by fitting state-of-the art spectral models to the Balmer lines 
\citep{tremblay09}, and calculating both atmospheric parameters (e.g., log($g$), 
$T_{\rm eff}$, and cooling age) and theoretical luminosities.  \cite{tremblay12} 
also simulated the field contamination along this sightline in their analysis.  By 
comparing the spectroscopic and kinematic distances, as well as the cooling ages of the 
new stars to the cluster age, they confirmed five of the new 
candidates as likely members of the Hyades.\footnote{Hyades white dwarf WD0231-054 is excluded 
since the photometric temperature doesn't agree with the spectroscopic temperature.}  
The other candidates are not explicitly excluded from membership.  Radial velocities of several of 
these candidates were also observed by \cite{zuckerman13}, who confirm three of 
the new candidates as bona-fide members of the Hyades, but also reject WD0743+442.  The final 
list of Hyades members that we consider, including these new stars and the seven classical 
members that are not in binaries, is presented in Table~\ref{table1}.  
The atmospheric parameters for these stars have been taken directly from \cite{tremblay12}.

For the Praesepe, earlier studies measured five white dwarf candidates 
\citep{luyten62,eggen65,anthony-twarog82,anthony-twarog84,claver01}, and more recent 
observations have identified an additional six white dwarf candidates 
\citep{dobbie04,dobbie06}.  \cite{casewell09} present a careful examination of nine of 
these stars, based on high-resolution optical spectroscopy, and show contamination in 
the sample from a magnetic white dwarf and a likely field white dwarf.  Their final 
sample of Praesepe white dwarf members includes seven white dwarfs, which are listed in 
Table~\ref{table1} (although, see below for the atmospheric properties of these stars).


\section{Initial and Final Masses} \label{sec:masses}

The atmospheric properties of the 4 white dwarfs in NGC~6819 and NGC~7789, and the 18 white dwarfs 
in the Hyades and Praesepe clusters are listed in Table~\ref{table1}.  These properties, including the white 
dwarf masses ($M_{\rm final}$), were calculated by \cite{kalirai09}, \cite{tremblay12} and 
\cite{casewell09} using the successful technique of fitting the Balmer lines in the spectra with 
model atmospheres \citep{bergeron92}.  As has been demonstrated numerous times, this technique 
leads to accurate parameters provided the spectra have high signal-to-noise ratio and sensitivity 
to the higher order Balmer lines \citep{kleinman13}.  However, both the \cite{kalirai09} and \cite{casewell09} 
studies modeled the white dwarf spectra using older line profiles compared to those presented 
in \cite{tremblay09}.  We therefore apply a small correction to these results to place them on 
the same foundation as the new Hyades measurements (i.e., from Figure 12 in Tremblay \& Bergeron 
2009).  For example, for the Praesepe white dwarfs, this correction is $+$400~K in $T_{\rm eff}$ 
and $+$0.1~dex in log($g$).  All of the updated atmospheric properties, including the final masses 
of the white dwarfs, are presented in Table~\ref{table1}.

Progenitor masses for these white dwarfs can be calculated by taking advantage of their membership 
in the four clusters (e.g., see Kalirai et~al.\ 2005 for a similar study in another intermediate age 
cluster).  First, the mass and temperature of each white dwarf uniquely sets its cooling age 
($t_{\rm cool}$), which represents the time since that 
white dwarf left the tip of the AGB.  By subtracting this cooling age from the age of the 
star cluster, we arrive at the lifetime of the progenitor star that made the white dwarf 
(i.e., the dominant main-sequence lifetime plus the post main-sequence lifetime up to the 
tip of the AGB).  The ages of the clusters are taken from earlier studies -- 2.5 Gyr for NGC~6819 
and 1.4~Gyr for NGC~7789 \citep{kalirai01,kalirai08}, and 625~Myr for the Hyades and Praesepe 
\citep{perryman98,claver01}.  The progenitor masses of the stars 
($M_{\rm initial}$) follow from standard stellar models at the cluster metallicity, and are listed in 
the last column of Table~\ref{table1} (new Bressan et~al.\ 2012 models).  The sensitivity of 
these initial masses to mild changes in the metallicity or age of the star clusters is small.  
For example, a shift in the age from the default 625~Myr by $\pm$50~Myr leads to initial masses 
that are $<$3\% smaller or larger, and a change in the metallicity of $\Delta$Z = 0.05 leads 
to a similar effect on the masses.  Such effects on the ages of the older clusters NGC~6819 and NGC~7789 
lead to even smaller uncertainties.

The individual measurements for stars in each cluster presented in Table~\ref{table1} are averaged into four 
initial-final mass pairs in Table~\ref{table2}.  Also included is the resulting integrated mass loss through 
stellar evolution.  For $M_{\rm initial}$ $\sim$ 3~$M_\odot$, our results demonstrate that stars will 
lose 75\% of their mass to the interstellar medium.  As expected, the mass loss is measured to be very 
similar for the Hyades and Praesepe clusters, given their identical age and metallicity.


\section{Core-Mass Growth on the TP-AGB} \label{sec:coremassgrowth}

The core mass at the first thermal pulse, $M_{\rm c, 1tp}$, is
primarily a function of initial stellar mass and chemical composition.
A general agreement exists among different stellar evolution models on
the trend of $M_{\rm c, 1tp}$ with the stellar mass. For instance, a
minimum of $M_{\rm c, 1tp}$ is expected in correspondence to the
maximum mass, $M_{\rm He-F}$, for a star to develop an electron
degenerate He-core, while the first occurrence of the second dredge-up 
in intermediate-mass ($M_{\rm initial} \simeq 3-4 M_{\odot}$) 
produces a change in the slope (inflection point) of the 
$M_{\rm c, 1tp}-M_{\rm initial}$ relation that runs flatter at higher masses.
Clearly, precise predictions of these features do depend on
the physics adopted in stellar models (see e.g., Wagenhuber \&
Groenewegen 1998). However, the current theoretical dispersion 
in $M_{\rm c, 1tp}$ is much smaller than the uncertainties 
in the final masses due to the uncertainties in the subsequent 
TP-AGB evolution.  In this sense, $M_{\rm c, 1tp}$ may be considered 
a robust prediction of stellar models. 

We take $M_{\rm c, 1tp}$ from the new stellar evolutionary models in \cite{bressan12} (i.e., 
the PARSEC code: PAdova \& TRieste Stellar Evolution Code) for initial composition 
$Z_{\rm initial}=0.02,\, Y_{\rm i}=0.284$,  with a scaled-solar distribution 
of metal abundances according to Caffau et~al.\ (2011), -- this corresponds to 
Solar metallicity $Z_{\odot}=0.01524$.  For example, over the range of initial 
masses spanned by the Hyades and Praesepe white dwarfs in Table~1, $M_{\rm{ c, 1tp}}$ = 0.60~$M_\odot$ at 
$M_{\rm initial}$ = 2.8~$M_\odot$, $M_{\rm{ c, 1tp}}$ = 0.70~$M_\odot$ at $M_{\rm initial}$ = 3.3~$M_\odot$, 
and $M_{\rm{ c, 1tp}}$ = 0.76~$M_\odot$ at $M_{\rm initial}$ = 3.8~$M_\odot$.  


\begin{table*}
\begin{center}
\caption{Summary of Initial-Final Mass Pairs for Each Cluster}
\begin{tabular}{cccccc}
\hline
\hline
\multicolumn{1}{c}{Cluster} &
\multicolumn{1}{c}{$M_{\rm initial}$ ($M_\odot$)} & \multicolumn{1}{c}{$M_{\rm final}$ ($M_\odot$)} &
\multicolumn{1}{c}{Integrated Mass Loss Through Post-} \\

\multicolumn{1}{c}{} &
\multicolumn{1}{c}{} & \multicolumn{1}{c}{} &
\multicolumn{1}{c}{Main Sequence Evolution (\%)} \\

\hline
NGC~6819  &  1.61 $\pm$ 0.01  &  0.575 $\pm$ 0.015 & 64.3  \\
NGC~7789  &  2.02 $\pm$ 0.00  &  0.650 $\pm$ 0.010 & 67.8  \\
Hyades    &  3.22 $\pm$ 0.11  &  0.749 $\pm$ 0.017 & 76.8  \\
Praesepe  &  3.56 $\pm$ 0.10  &  0.837 $\pm$ 0.020 & 76.5  \\
\hline
\end{tabular}
\label{table2}
\end{center}
\end{table*}


Beyond the first thermal pulse, the subsequent TP-AGB is challenging to model because of the complex 
interplay of many physical processes, which are often affected by severe uncertainties.  During 
this phase, the mass of the H-exhausted core increases following the outward advancement of the 
H-burning shell during the quiescent inter-pulse periods, while the mass may be temporarily reduced 
at each third dredge-up event, by an amount that depends on the depth of the envelope penetration.
In the meantime the stellar envelope is progressively lost by stellar winds.  Therefore, the size 
of stellar core increase is controlled by the competition between (a) the speed of 
displacement of the H-burning shell, that fixes the core growth rate, (b) 
the strength of mass loss, that determines  the TP-AGB timescale, and (c) 
the efficiency of the third dredge-up (if it occurs), that lessens the effective mass increment. 
While the former aspect mainly relies on well-established properties of nuclear reactions,  
the latter two processes, i.e., mass loss and third dredge-up,  are still not robustly 
assessed on theoretical grounds.  For more information, see \cite{marigogirardi_01,marigo13b}.

The end product of the TP-AGB is the nuclear-processed core, the C-O white dwarf.  The masses of the 
22 white dwarfs in Table~\ref{table1} therefore provide a novel method to directly measure the core 
growth on the TP-AGB, $\Delta M_{\rm growth}$ = $M_{\rm final}$ $-$ $M_{\rm{ c, 1tp}}$.  We 
illustrate this in Figure~\ref{fig:growth}, both for the individual raw data (open circles with 
error bars) and five (straight) average values across the initial mass spectrum.  The averages are 
calculated by treating each of the NGC~6819 and NGC~7819 pairs separately, and then defining three 
mass bins between 2.5 $<$ $M$ $<$ 4.0~$M_\odot$ with bin width 0.5~$M_\odot$ for the 18 Hyades and 
Praesepe white dwarfs.  The averages are shown as darker filled circles and connected with a thick 
black line.  The uncertainties in these values are the errors in the mean for each average.  The 
core-mass growth is shown as a percentage, 
$\Delta M_{\rm growth}$/$M_{\rm{ c, 1tp}}$.  The binned averages illustrates that $\Delta M_{\rm growth}$ 
increases rapidly from 10\% to 30\% for stars with $M_{\rm initial}$ = 1.6 to 2.0~$M_\odot$, and at larger 
masses decreases down to $\sim$10\% at $M_{\rm initial}$ = 3.4~$M_\odot$.  There is a small hint of an upturn at 
larger masses, suggesting that the core-mass growth is $\gtrsim$10\% up to $M_{\rm initial}$ = 3.8~$M_\odot$.

For $M_{\rm initial}$ $>$ 3~$M_\odot$, our results are systematically lower than those reported in the 
similar study by \cite{bird11}, by as much as a factor of two.  Although their study also looked at white dwarfs 
in the Hyades and Praesepe clusters (not including the new discoveries and uniform measurements from 
the Tremblay et~al.\ 2012 models), they also included white dwarf measurements in two other star clusters over 
this mass range (i.e., with different ages and metallicities).  


\section{Testing TP-AGB Models} \label{sec_models}

The measurement of $\Delta M_{\rm growth}$ over $M_{\rm initial}$ = 1.6 -- 3.8~$M_\odot$ 
provides a new test to the latest evolutionary models of TP-AGB stars.  New calculations 
by \cite{marigo13} offer significant advances over previous generation models.  These 
models begin at the first thermal pulse, extracted from the new \cite{bressan12} stellar 
models, and continue to the complete ejection of the envelope due to winds \citep{marigo08}.  
Compared to past releases \citep{marigogirardi_07, girardi10} the new tracks now include a 
more accurate treatment of the star's energetics (the core mass-luminosity relation and 
its break-down due to hot-bottom burning are self-consistently predicted), and rely on the 
first ever on-the-fly computations of detailed molecular chemistry and gas opacities 
\citep{marigo09}.  This new advance guarantees full consistency between the envelope structure 
and the surface chemical abundances, and therefore robustly tracks the impact of third 
dredge-up episodes and hot-bottom burning.

In this work, we explore the dependence of the predicted final mass left at the end of the 
TP-AGB phase to 1.) the efficiency of the third dredge-up, and 2.) the mass loss, starting 
from a reference set of TP-AGB models, as described in Marigo et al. (2013).  The 
occurrence of the third dredge-up is determined with the aid of envelope integrations
at the stage of the post-flash luminosity peak, checking if the condition 
$T_{\rm bce} > T_{\rm dred}$ is fulfilled, i.e., the temperature at the base of the 
convective envelope exceeds a minimum value \citep[more details in][]{marigo13}.  
For the present calculations we set $\log(T_{\rm dred}) = 6.6$, a value somewhat 
larger than the $\log(T_{\rm dred}) = 6.4$ that was assumed for the test models 
presented in Marigo et~al.\ (2013).  Increasing $T_{\rm dred}$ causes a later onset 
of the third dredge-up, i.e., at larger core masses, which is a more suitable choice 
for describing the formation of carbon stars at higher metallicities, as suggested by 
previous full model calculations \citep{karakas02} and calibration studies 
(Marigo et~al.\ 1999).

The efficiency\footnote{The efficiency of the third dredge-up is usually expressed 
with $\lambda=\frac{\Delta M_{\rm dup}}{\Delta M_{\rm c}}$, defined as the fraction of 
the core mass increment over an inter-pulse period ($\Delta M_{\rm c}$), that is dredged-up 
to the surface at the next thermal pulse (with mass $\Delta M_{\rm dup}$)} of the third 
dredge-up $\lambda$, as a function of stellar mass  and metallicity, is taken from the 
relations of \citet[][hereafter also K02]{karakas02}, that fit the results of their full 
TP-AGB models.  The K02 formalism represents our initial prescription for the third dredge-up,
which will be then varied to explore the sensitivity of the predicted final masses 
to different efficiencies of the mixing episodes, and to eventually obtain calibrated 
relations for $\lambda$ as a function of the stellar mass.  The mass loss prescription is 
similar to that adopted in Girardi et~al.\ (2010).  The Reimers mass loss formulation 
with an efficiency parameter $\eta=0.2$ \citep[following the recent asteroseismologic 
calibration of][]{Miglio_etal12} is assumed in the initial stages, followed by an 
exponentially increasing mass-loss rate relation, derived from computations of 
periodically-shocked dusty atmospheres \citep{Bedijn_88}.
 
Similar to other descriptions, the Marigo et~al.\ (2013) models take the efficiencies 
of both the third dredge-up and mass loss as free parameters, to be calibrated with 
observations.  Indeed, the initial-final mass relation provides us with an important tool 
to put constraints on these two processes. In this perspective, besides the default choice 
of parameters, we consider several additional prescriptions for both processes.  
Given its flexibility, physical accuracy, and fast performance, the {\sc colibri} code 
developed by \citet{marigo13} is an appropriate tool to carry out extensive 
exploration and calibration analyses.


\subsection{Characterizing the Significance of the \\ Third Dredge-Up} \label{sec_3dup}

The third dredge-up affects the core-mass growth on the TP-AGB in two main modes. 

The first effect is the {\em direct} reduction of its mass: every time a dredge-up 
episode takes place with an efficiency $\lambda$, the core mass is almost instantaneously 
turned down by an amount $\lambda\Delta M_c$.  Unfortunately, the efficiency $\lambda$ 
is one of the most uncertain parameters of TP-AGB star modeling as it is 
found to vary significantly from study to study, depending on the adopted treatment 
of convection, mixing, and numerics (see e.g., Marigo 2012 for a review).

An {\em indirect} effect is driven by the changes in the surface chemical composition 
caused by the penetration of the base of the convective envelope into the inter-shell region. 
In fact, each dredge-up event results in a mixing of material (mainly $^4$He, $^{12}$C, 
$^{22}$Ne, Na, Mg and Al isotopes, and slow-neutron capture elements) left by the 
pulse-driven convective zone to the outer layers. In particular, the enrichment 
in primary carbon causes the surface C/O ratio to increase.  As soon as the number of 
carbon atoms exceeds that of oxygen (i.e., C/O$>1$) an abrupt change in the molecular 
equilibria causes a sudden rise of the atmospheric opacity \citep{marigo02}.  
In turn, this results in lower  effective temperatures  and increased mass loss 
from dust-driven winds \citep{marigogirardi_07, mattsson_etal10}.  As a consequence, 
the TP-AGB lifetime is shorter and the growth of the core mass is smaller than otherwise 
predicted neglecting the enhancement of the carbon-bearing opacity.  

\begin{figure}[ht]
\leavevmode 
\resizebox{\hsize}{!}{\includegraphics{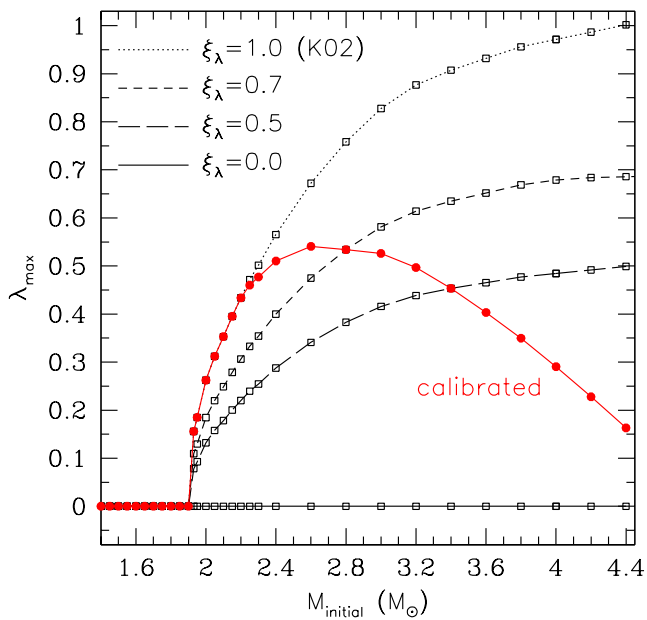}}
\caption{Maximum efficiency, $\lambda_{\rm max}$, of the third dredge-up attained 
during the TP-AGB evolution as a function of the initial stellar mass.  The four curves 
correspond to selected values of the variation factor $\xi_{\lambda}$, defined by the relation 
$\lambda_{\rm max}= \xi_{\lambda} \, \lambda_{\rm max}^{\rm K02}$, where 
$\lambda_{\rm max}^{\rm K02}$ denotes the reference predictions of \citet[][K02]{karakas02}.  
Note that $\xi_{\lambda}=0$ refers to models without third dredge-up. The calibrated relation 
(red line connecting filled circles), based on the new observed average core-mass 
growth from our data, exhibits a non-monotonic behavior with the stellar mass.}
\label{fig_lmax}
\end{figure}

According to the K02 models, $\lambda$ quickly increases from one thermal pulse to another 
until it reaches a maximum, $\lambda_{\rm max}$, whose value typically increases with the 
stellar mass, while it decreases at larger metallicity. To explore the effect of the third 
dredge-up, we vary its efficiency  by simply multiplying the original K02 $\lambda_{\rm max}$ 
by four selected factors, i.e., $\lambda_{\rm max} = \xi_{\lambda} 
\lambda_{\rm max}^{\rm K02}$, with $\xi_{\lambda} =0.0, 0.5, 0.8, 1.0$, as shown in 
Figure~\ref{fig_lmax}.  They represent a sequence of increasing efficiency of the 
third dredge-up, starting from no dredge-up   ($\xi_{\lambda} =0.0)$, up to recover 
the reference K02 relations ($\xi_{\lambda} =1.0)$. Since this latter case
yields already rather large efficiencies ($\lambda_{\rm max} \ga 0.8-0.9$)
for intermediate-mass stars ($M_{\rm initial}> 2.5 \, M_{\odot}$), 
we do not consider larger value, i.e., $\xi_{\lambda} >1.0$.

\begin{figure*}[ht]
\begin{center}
\leavevmode 
\includegraphics[width=12.0cm]{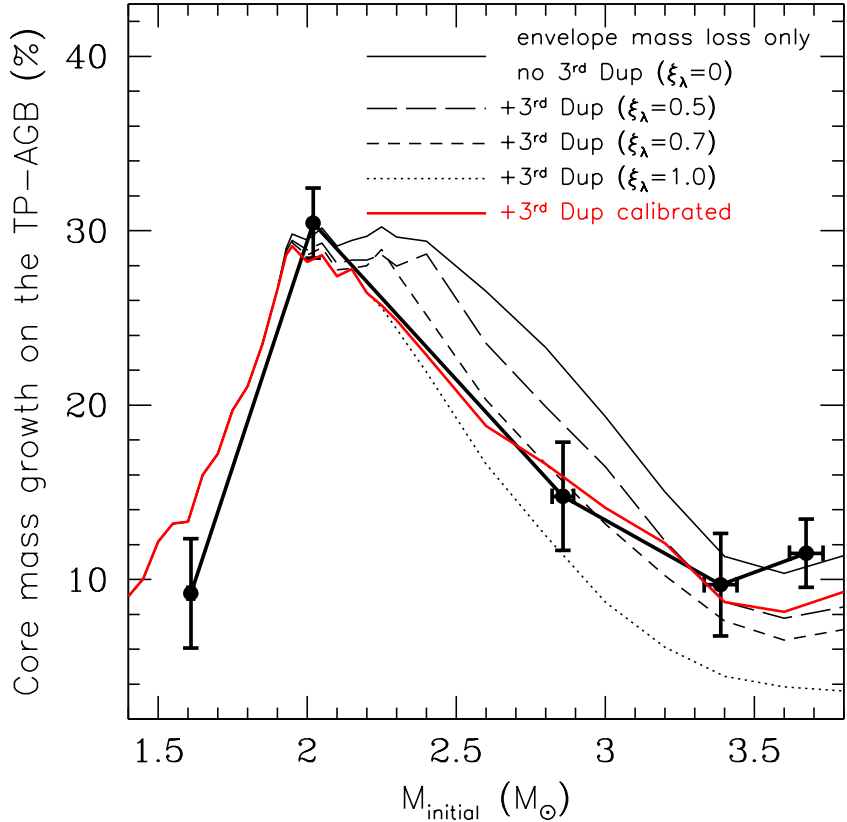}
\end{center}
\vspace{-0.4cm}
\caption{Our measurements for the growth of the stellar core on the TP-AGB is shown as a darker 
line with filled circles, and compared to five theoretical models of the TP-AGB phase of stellar 
evolution from \cite{marigo13}.  Each of these models only differs in the treatment of the efficiency 
of the third dredge-up process, as described in \S~\ref{sec_3dup}.  The general agreement between 
these models and the new data is excellent.  Within the set of models, the short dashed curve 
representing a parametrization of $\lambda_{\rm max}$ = 0.7$\lambda_{\rm max}^{\rm K02}$ for 
the efficiency of the third dredge-up is able to recover the data very well.  A refined 
agreement is obtained with an empirical calibration of the third dredge-up efficiency as a 
function of the initial stellar mass (red line).  Our observations therefore suggest that the third 
dredge-up does play a role in governing the growth of the core on the TP-AGB, however we will see 
later in \S~\ref{sec_mloss} that it is not the dominant effect.}
\label{fig_3dup}
\end{figure*}


\begin{figure*}[ht]
\begin{minipage}{0.49\textwidth}
\resizebox{1.0\hsize}{!}{\includegraphics{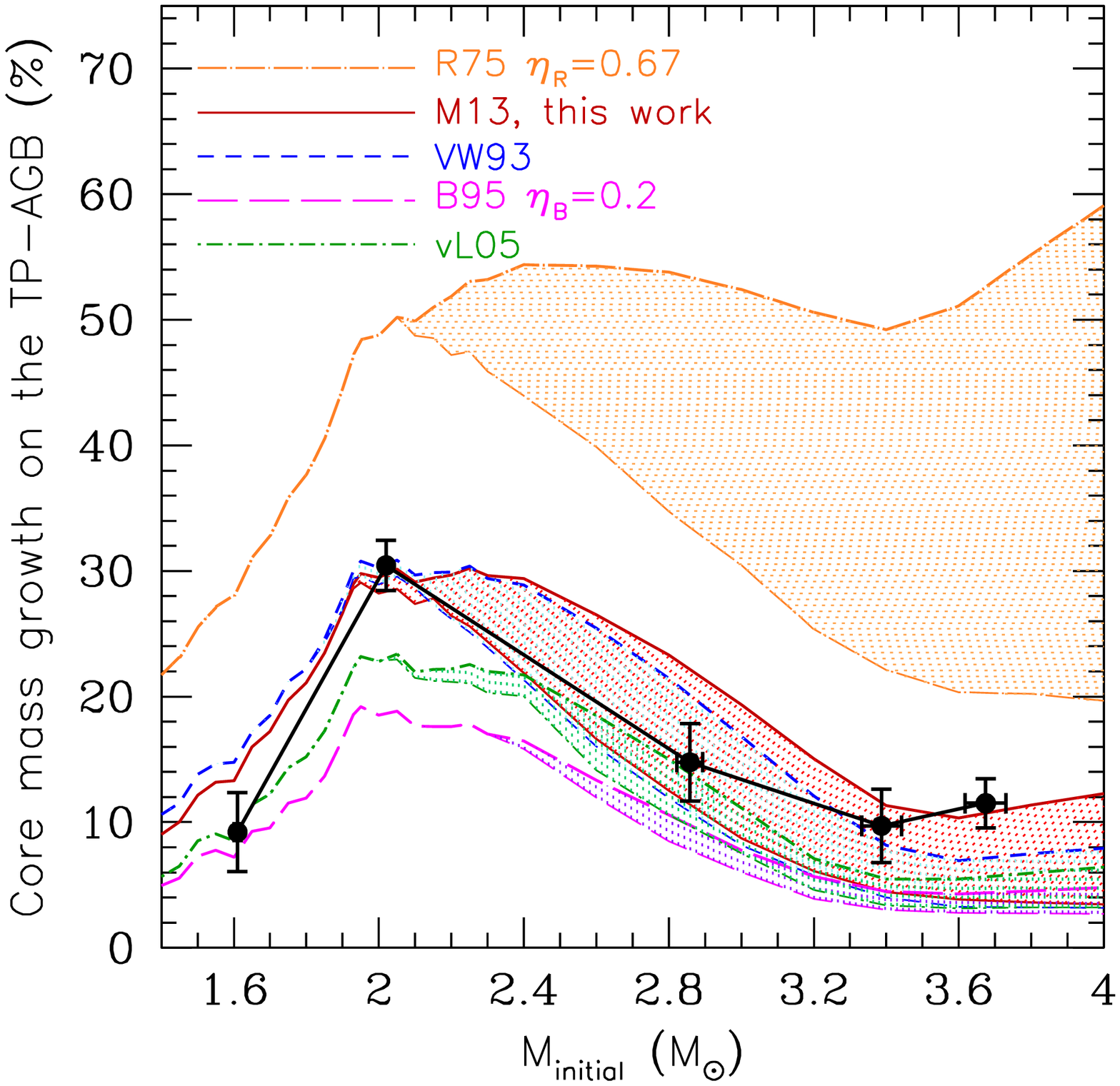}}
\end{minipage}
\hfill
\begin{minipage}{0.49\textwidth}
\resizebox{1.0\hsize}{!}{\includegraphics{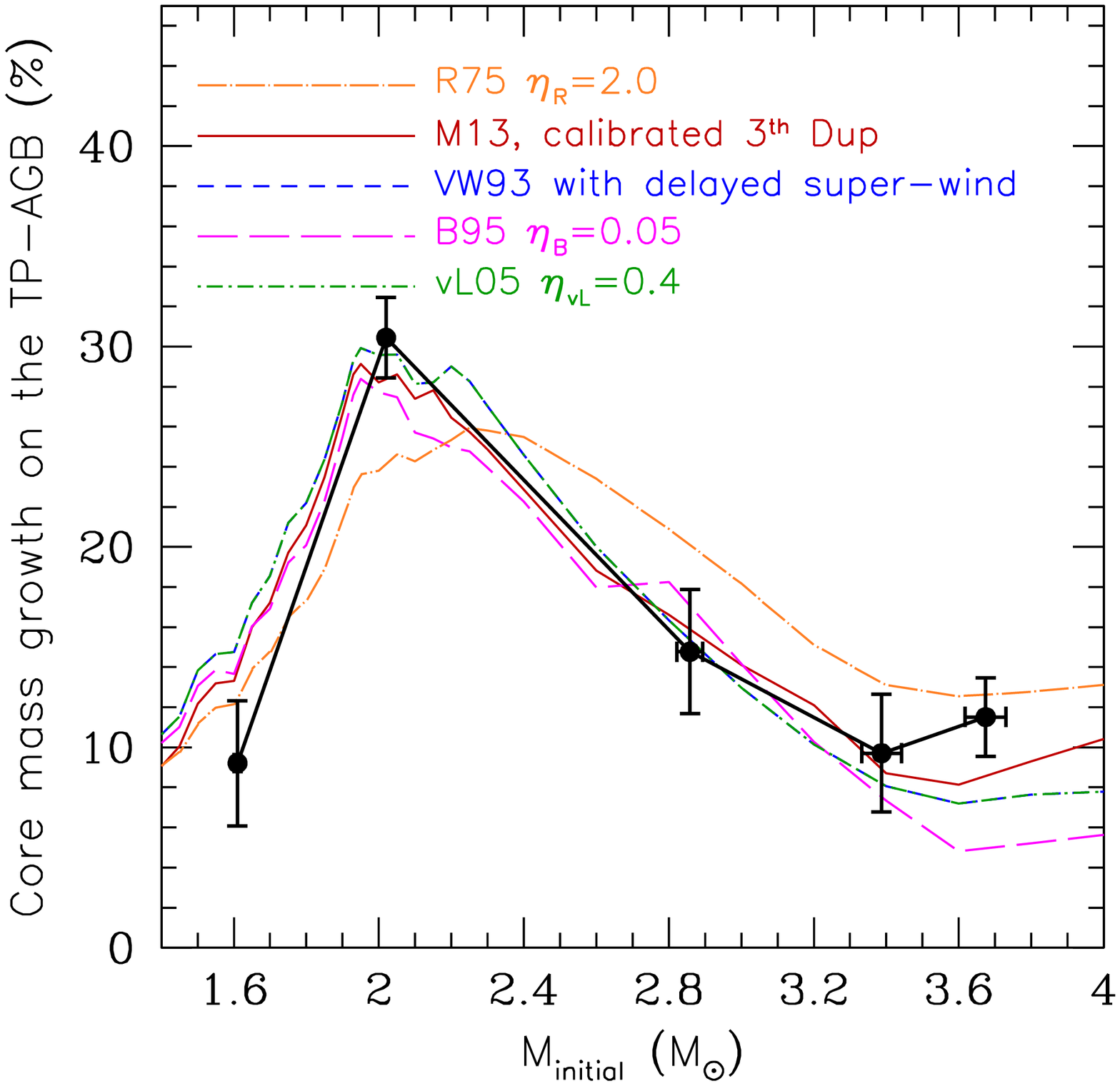}}
\end{minipage}
\caption{The same as in Figure~\ref{fig_3dup}, but showing the predictions with five 
different descriptions for mass loss on the TP-AGB phase, namely: the Reimers law (R75, 
orange curves), our reference prescription (Marigo et~al.\ 2013; red curves), the \citet[][VW93, 
blue curves]{VassiliadisWood_93}, the \citet[][B95, magenta curves]{Blocker95}, and the 
\citet[][vL05, green curves]{vanLoon_etal05}.  {\bf Left panel}: For each mass-loss case, 
the hatched region encompasses the range of core mass growth expected when varying the 
third dredge-up  efficiency between two extremes, namely: $\xi_{\lambda}=1$ (the original 
K02 prescription, thin line) and $\xi_{\lambda}=0$ (no dredge-up, thick line).
The latter case corresponds to the maximum growth of the core mass 
allowed by the corresponding mass-loss relation.  {\bf Right panel}: Results obtained 
with modified versions of the same mass-loss prescriptions (except for the Marigo et~al.\ 2013 
case), adopting suitable efficiency parameters or revised relations so as to approach the 
observational constraints in our study.}
\label{fig_mloss}
\end{figure*}


Based on the four curves in Figure~\ref{fig_lmax}, we calculate theoretical 
sequences for TP-AGB evolution, and illustrate the resulting core mass growth in 
Figure~\ref{fig_3dup}. All of the predictions have the same shape and approximate 
normalization as the new observations (darker line with filled circles).  This agreement 
is a remarkable validation of these models at $Z_{\rm initial}$ = 0.02, which lack 
strong observational tests.  The top solid black curve predicts that the maximum 
core-mass growth in the absence of any third dredge-up reaches $\Delta M_{\rm growth}$ 
= 30\% at $M_{\rm initial}$ $\sim$ 2~$M_\odot$, decreasing steadily to 
$\Delta M_{\rm growth}$ = 23\% at $M_{\rm initial}$ = 2.8~$M_\odot$ and 
$\Delta M_{\rm growth}$ = 11\% at $M_{\rm initial}$ = 3.8~$M_\odot$.  

We note that for $M_{\rm initial}$ $\la$ 1.9 $M_{\odot}$ all the curves coincide, since at
these masses and for $Z_{\rm initial}=0.02$ the third dredge-up is expected not to take place. 
At larger masses, $M_{\rm initial}$ $\ga$ 1.9 $M_{\odot}$, the curves start to deviate as a 
consequence of the third dredge-up.  The three sequences below the top-most model 
($\xi_{\lambda}=0$) each correspond to the same mass-loss law (Bedijn 1988, 
see Section~\ref{sec_mloss}), but with increasing efficiency 
of the third dredge-up process (as indicated in the label).  These models progressively 
predict a smaller growth in the stellar core, as expected given the direct reduction of 
the H-exhausted core following each third dredge-up event and the shorter lifetime of 
the TP-AGB phase.  

It follows that, for a given mass-loss prescription, the measurement of the core-mass 
growth from the white dwarfs is helpful to constrain the third dredge-up as a function 
of the progenitor's stellar mass, hence of the age.  Based on the observed average 
relationship shown in Figure~\ref{fig_3dup}, we have tentatively calibrated the dredge-up 
parameter $\lambda_{\rm max}$ as a function of the stellar mass, so as to obtain the 
best match with the data using our reference mass-loss prescription.  The corresponding 
$\lambda_{\rm max}(M_{\rm initial})$ relation is plotted in Figure~\ref{fig_lmax} as a 
red curve.

A few interesting implications can be drawn.  First, at metallicity  $Z_{\rm initial}=0.02$ -- 
corresponding to [Fe/H]$\simeq 0.1$ for the adopted solar mixture -- the third dredge-up 
would occur only in stars with $M_{\rm initial}$ $\ga$ 2$M_{\odot}$. Second, in the range 
2$M_{\odot}$ $\la M_{\rm initial}$ $\la$ 3.0$M_{\odot}$, its maximum efficiency should 
increase with the stellar mass from zero up to $\lambda_{\rm max}\approx 0.5$ (see 
Figure~\ref{fig_lmax}).  Third, the data seem to suggest that at larger masses, 
$M_{\rm initial}$ $>$ 3.0$M_{\odot}$, the third dredge-up should become progressively 
less efficient, with $\lambda_{\rm max}$ declining towards low values.  The decreasing 
trend of $\lambda_{\rm max}$ is required to recover the rising trend in the growth of 
the core mass that the Praesepe cluster white dwarfs seem to suggest.  We should note 
that this indication is at variance with the K02 models, that instead predict larger 
values for $\lambda_{\rm max}$ at increasing stellar mass.  Further investigation on 
both theoretical and observational grounds is deserved before a conclusion on this
aspect can be drawn. Clearly, this may have important implications for the chemical 
yields produced by more massive TP-AGB stars.

In summary, with the present prescription for the third dredge-up, we expect a modest 
carbon star formation at metallicity $Z_{\rm initial} \simeq 0.02$, mostly confined in 
stars with masses 2~$M_{\odot}$ $\la M_{\rm initial}$ $\la$ 3~$M_{\odot}$.  The corresponding 
final surface C/O ratios remain quite low, 1 $<$ C/O $\lesssim$ 1.3 (last column of 
Table~\ref{table3}), and the fraction of the TP-AGB lifetime spent in the C-star mode 
reaches a maximum of $\simeq 23\%$ at  $M_{\rm initial}\simeq 2.6\,M_{\odot}$.
It is interesting to notice that this result is  nicely supported
by the recent study of   \citet{Boyer_etal13}, that  has revealed a dramatic 
scarcity of carbon stars in the inner disk of Andromeda galaxy, characterized 
by a high metallicity (i.e., [Fe/H] $\simeq$ $+$0.1), comparable to that 
considered here.

Finally, we plot in Figure~\ref{fig_3dup} the theoretical curve for the
core-mass growth, obtained with our calibrated function for $\lambda_{\rm max}$.  
Our best-fit model shows consistency within $\sim$2\% at all masses. 


\subsection{Characterizing the Significance of Mass Loss} \label{sec_mloss}

We investigate the influence of stellar winds in controlling the growth of the core mass 
by running the same set of TP-AGB models for initial metallicity $Z_{\rm initial}=0.02$, 
but adopting four additional options for the mass-loss rates, namely:
the classical \citet[][also R75]{Reimers75} law, and the popular formulas
of \citet[][also VW93]{VassiliadisWood_93}, \citet[][also B95]{Blocker95},  
and \citet[][also vL05]{vanLoon_etal05}.

Though the Reimers law is known to be inadequate to describe the evolution of 
the mass-loss rates along the TP-AGB \citep{Blocker95, SchroederCuntz_05, Groen_etal09, Cranmer_11}, 
it is still a classical reference in many studies and its behavior was taken 
into account, for instance, to infer the metallicity dependence of the 
TP-AGB fuel in the stellar population synthesis models of \citet{maraston05}. 
In that work, the author concluded that the TP-AGB fuel as a function of age, 
calibrated on Magellanic Clouds clusters, would correspond to adopting 
the Reimers law with $\eta_{\rm R}=2/3$ in TP-AGB calculations 
\citep{RenziniVoli_81}.  This value represents quite a low efficiency compared 
to $\eta_{\rm R}=5$ as derived by \citet{GroenJong_93} to reproduce the observed 
AGB star luminosity functions in the LMC.  It is therefore interesting to check 
the Reimers assumption with our new TP-AGB models and the new white dwarf data. 

The \citet{VassiliadisWood_93} model, calibrated on the empirical relation 
between mass-loss rates and pulsation periods of variable AGB stars, has become 
a reference recipe to describe mass loss during the AGB.  As a first approach, we 
adopt the original formulation (equations 1 and 2 of VW93).

The \citet{Blocker95} relation is also a popular prescription in present-day 
TP-AGB models, and is characterized by quite a steep luminosity dependence.
Following the indications of the original paper of  \citet{Blocker95}, 
we initially assume the Reimers law with an efficiency parameter $\eta_{\rm R}=0.2$
and, as soon as the pulsation period in the fundamental mode exceeds 100 days, we then 
switch to the B95 formula keeping the same efficiency parameter, 
$\eta_{\rm B}=0.2$.

Finally, we test the relation derived by \citet{vanLoon_etal05} on the basis of 
spectroscopic  and photometric observations of dust-enshrouded red giants in the LMC.
Similarly to the other cases, we first adopt the the Reimers law with $\eta_{\rm R}=0.2$, 
and  then we activate the vL05 formula as the pulsation periods becomes longer than 300 days
($\simeq$ to the minimum period of the stars in vL05 calibration sample).

For each mass-loss prescription, we consider two choices of the third dredge-up efficiency, 
namely:  $\xi_{\lambda}=1$, that is  the standard case $\lambda^{\rm K02}_{\rm max}$ predicted 
by \citet{karakas02}, and $\xi_{\lambda}=0$, that corresponds to the absence of any dredge-up 
event.  In this way, for each mass-loss law, we can sample the characteristic dispersion in 
the core mass growth that derives by variations in the third dredge-up efficiency.  In particular, 
the case of $\xi_{\lambda}=0$ yields the upper limit of the core mass increment attainable with 
a given mass-loss prescription.  The results are shown in Figure~\ref{fig_mloss}.  First, we 
note that the range of the core mass growth enclosed between $\xi_{\lambda}=0$ and $\xi_{\lambda}=1$
(hatched areas in Figure~\ref{fig_mloss}) anti-correlates with the average efficiency of the 
mass loss, being quite narrow with the B95 and vL05 relations, while becoming much wider with 
the R75 law. 

The final masses obtained with the \citet{VassiliadisWood_93} formalism compare with 
the data very well, and are strikingly close to those derived with our reference 
mass-loss prescription, which is based on the \citet{Bedijn_88} formalism.  
We recall that both relations are empirically calibrated, but the calibration samples 
of AGB stars and the measured quantities are different, i.e., pulsation periods for VW93; 
radii, masses and effective temperatures for the Bedijn (1988)-like method. The convergence of the 
predictions, and the nice agreement with the new observations, at least for the 
metallicity under consideration, is a promising step towards a more robust AGB calibration. 

In this context, the comparison with the observed core mass growth allows one to reject 
unsuitable mass-loss efficiencies.  For instance, the core-mass increment on the TP-AGB 
obtained with the \citet{Blocker95} relation and $\eta_{\rm B}=0.2$ is always too small, 
even invoking the most favorable case of no third dredge-up.  The same seems to apply, 
though to a lesser extent, also to the \citet{vanLoon_etal05} empirical relation.  The 
adopted B95 and vL05 mass-loss formulations do not allow the core to grow enough on the 
TP-AGB, at least for the case of slightly super-solar initial metallicity, 
$Z_{\rm initial} \simeq 0.02$ (or equivalently, [Fe/H] $\simeq$ 0.1).  We note that the 
high efficiencies of the B95 and L05 mass-loss relations are related to different 
functional dependences.  While the strength of the B95 relation is mostly controlled
by the increase in luminosity ($\dot M_{\rm B95} \propto L^{4.2}$), hence being
particularly efficient in more massive AGB experiencing HBB, the intensity of the 
vL05  mass loss is dictated by the steep sensitivity to the effective temperature 
($\dot M_{\rm vL05} \propto T_{\rm eff}^{-6.3}$), so that it is expected to affect 
particularly TP-AGB models of higher metallicities, like those considered in this 
work.

Contrary to the B95 and vL05 mass-loss rates, the opposite problem arises with the Reimers 
law adopting $\eta_{\rm R}=2/3$:  the predicted mass loss is too weak, leading to an overestimate 
of the growth of the core, unless one were to assume that the efficiency of the third dredge-up 
remains close to unity for most of the TP-AGB evolution at any stellar mass.  As a trial, we have 
considered the case of an extremely strong third dredge-up, taking a high value of the 
multiplicative factor for the maximum efficiency, $\xi_{\lambda}=1.5$, and forcing an earlier onset 
of the mixing events by setting a lower temperature parameter, $\log(T_{\rm dred})=6.3$.  We find 
that the increase of the core mass is now lower, but still too high compared to the observation 
by roughly $50 \%$ at any initial stellar mass.  Moreover, such a deep third dredge-up leads to an 
efficient carbon star formation and quite large surface C/O ratios, of up to $4-5$.  This prediction 
seems unrealistic considering that, instead, Galactic carbon stars normally exhibit C/O ratios 
of just over unity, in any case never exceeding $1.8-2.0$ \citep{Lambert_etal86, Ohnaka_etal00}.

Interestingly, some of these findings are in line with the claims of other studies 
derived from independent arguments. For instance, lower efficiencies for the \citet{Blocker95} 
relation have been adopted by \citet{Ventura_etal00} ($\eta_{\rm B}=0.01$) to reproduce the 
luminosity functions of Li-rich giants in the LMC.  More recently, \citet{Kamath_etal10} have 
found that the B95 anticipates the AGB termination at too faint luminosities in models aimed 
at reproducing observations of AGB stars in MC clusters. In a follow-up study 
\citet{Kamath_etal12} suggest that the observed luminosity of the AGB tip MC clusters can be 
correctly recovered assuming that the pulsation period at which the super-wind starts in the 
VW93 mass-loss prescription is delayed from $P \simeq$ 500 days to $P \simeq$ 700 -- 800 days.  
In this framework it is therefore useful to revise all of these mass-loss prescriptions and 
to find suitable values of their efficiency parameters, or to introduce other modifications 
that may improve the comparison with the observations.  

The right panel of Figure~\ref{fig_mloss} shows the results obtained by running additional sets 
of TP-AGB models with $Z_{\rm initial}=0.02$.  In all cases we adopt the relation for the 
third dredge-up efficiency corresponding to $\xi_{\lambda}=0.7$, while varying the 
mass-loss rates.  Specifically, we assume  the following set of parameters: $\eta_{\rm R}=2.0$ in the 
\citet{Reimers75} law;  $\eta_{\rm B}=0.05$ in the \citet{Blocker95} formula; inclusion of the 
multiplicative factor $\eta_{\rm vL}=0.4$ in the \citet{vanLoon_etal05} relation; delayed onset of 
the super-wind in the \citet{VassiliadisWood_93} prescription (their equation 3).

Postponing the super-wind in the VW93 mass-loss has the effect of slightly improving
the  comparison with the data towards larger stellar masses ($M_{\rm initial}$ $>$ 
$3.0\,M_{\odot}$), allowing a somewhat larger increase of the core mass.  For all of the 
other mass-loss laws, that instead suffered a more significant discrepancy (see left 
panel of Figure~\ref{fig_mloss}), the effect of adjusting the efficiency parameters 
is substantial, eventually leading to a satisfactory agreement with the observed data
in all cases (compare with right panel of Figure~\ref{fig_mloss}).  We also notice that 
the majority of the  mass-loss relations recover very well the morphology of the observed 
relation as a function of the initial stellar mass, predicting a peak at 
$M_{\rm initial}\simeq 2 \, M_{\odot}$ and declining wings at both lower and higher masses.  
The R75 law with $\eta_{\rm R}=2.0$ produces a somewhat worse trend, as the peak becomes
broader and shifted towards larger masses. 

As a final remark, we emphasize that the results in Figure~\ref{fig_mloss} show clearly 
that the main factor controlling the growth of the core mass in TP-AGB stellar models is 
the adopted mass-loss law.  The third dredge-up does play a non-negligible role but, in 
general, varying its efficiency produces  a narrower spread in the final masses than
that caused by assuming different mass-loss prescriptions, at least among those proposed 
in the literature for the TP-AGB phase.

\begin{figure}[ht]
\resizebox{1.0\hsize}{!}{\includegraphics{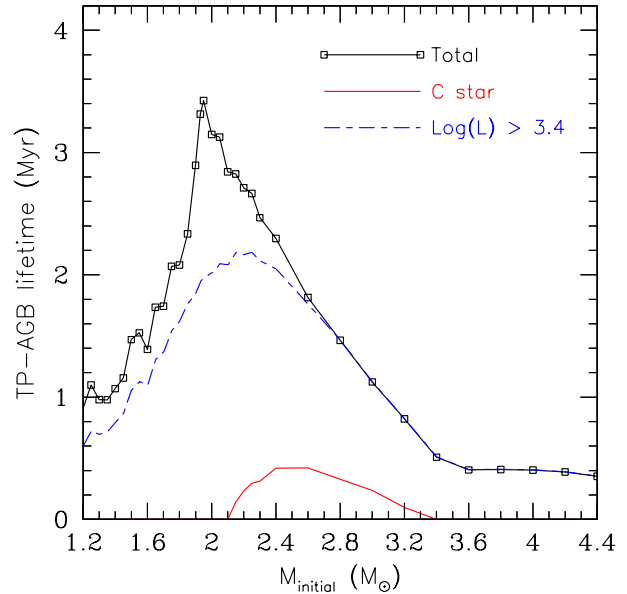}}
\caption{The lifetime of the TP-AGB phase from the Marigo et~al.\ (2013) models with 
initial metallicity $Z_{\rm initial} = 0.02$, and adopting a calibrated $\lambda_{\rm max}$ 
relation for the maximum efficiency of the third dredge-up.  The time spent at luminosities 
higher than the RGB tip, i.e., $\log(L/L_{\odot}) \simeq 3.4$, is also shown (blue dashed 
line), together with the C-star lifetime (red solid line).  The predicted TP-AGB core-mass 
growth in these models fits our new measurements very nicely, as demonstrated in 
\S~\ref{sec_3dup}.  At the peak core-mass growth in stars with $M_{\rm initial}$ 
$\sim$2~$M_\odot$, the lifetime of stars in the TP-AGB is $\tau$ $\sim$ 3.4~Myr, 
which reduces to $\sim$2~Myr if we consider the TP-AGB portion brighter than the RGB tip.  
For stars with $M_{\rm initial}$ $\sim$ 3~$M_\odot$, the  TP-AGB lifetime is 
$\tau$ $\sim$ 1~Myr, which drops to $\tau$ $\sim$ 0.45~Myr for $M_{\rm initial}$ 
$\sim$ 3.5~$M_\odot$.}
\label{fig:lifetime}
\end{figure}


\section{The Lifetime and Energy Output of \\ Stars on the TP-AGB} \label{sec:lum}

Given their luminous nature and the high level of mass loss suffered, the evolutionary 
properties of TP-AGB stars are critically important to establish meaningful constraints 
on the integrated light and chemical yields of stellar populations (e.g., we showed in 
\S\,\ref{sec:masses} that AGB stars with $\sim$3~$M_\odot$ will lose $\sim$75\% of their 
mass to the ISM).  For decades we have known that, owing to their high intrinsic 
brightness, TP-AGB stars contribute significantly to the total {\em bolometric} luminosity 
of single-burst stellar populations (SSP), reaching a maximum of about 40\% at ages from 
1 to 3 Gyr \citep{Frogel_etal90}.  It is worth noting that these classical estimates are 
actually quite uncertain and need to be revised, as recently demonstrated by \cite{girardi13}.  
The contribution of this phase to the near-IR luminosity 
may be as high as 80\% (see the review of Bruzual 2010, and also see Girardi \& Marigo 
2007 and Melbourne et~al.\ 2012).    

Presently, the treatment of the TP-AGB phase for evolutionary population synthesis models is 
disputed, leading to large uncertainties in the interpretation of astronomical 
observations.  For example, \cite{maraston06} fit the SEDs of high-redshift 
Spitzer galaxies, and demonstrate that the ages and masses are 60\% lower when adopting 
their TP-AGB models over the \cite{bruzual03} population synthesis models.  
On the other side, \citet{Kriek_etal10} show that the \citet{maraston05} models overpredict 
the rest-frame near-infrared luminosity of a sample of intermediate-redshift post-starburst 
galaxies.  More generally, \cite{conroy13} illustrates the strong degeneracy between the 
modeling of the TP-AGB phase of stellar evolution and the inferred metallicity, stellar 
mass, and star formation rate of galaxies.  The author stresses that the treatment of this 
phase is essential to avoid large systematic errors in galaxy properties.

A careful reconsideration of the TP-AGB phase, mainly in terms of its evolutionary properties 
as a function of age and metallicity, is therefore necessary at this stage.  Recently, 
\cite{girardi10} introduced a new way of calibrating the TP-AGB phase by directly comparing 
the number counts of AGB stars predicted on the color-magnitude diagram to that measured in 
a dozen nearby (low-metallicity) galaxies (from the {\it Hubble Space Telescope} 
ANGST/ANGRRR survey -- Dalcanton et~al.\ 2009).  The results show a dramatic improvement 
over the older models, both in terms of the TP-AGB tip luminosity and the general luminosity 
function.  The end product of this stellar evolution, with the new mass loss prescription 
(based on a Bedijn 1988-like formalism), suggests a white dwarf mass with $M_{\rm final}$ 
= 0.52 -- 0.54~$M_\odot$ for $M_{\rm initial}$ = 0.75 -- 0.85~$M_\odot$.  This prediction is 
in exact agreement with the measured remnant mass in the old, metal-poor globular cluster 
M4, $M_{\rm final}$ = 0.53 $\pm$ 0.01~$M_\odot$ \citep{kalirai09,kalirai12}.

\begin{figure}[]
\resizebox{1.0\hsize}{!}{\includegraphics{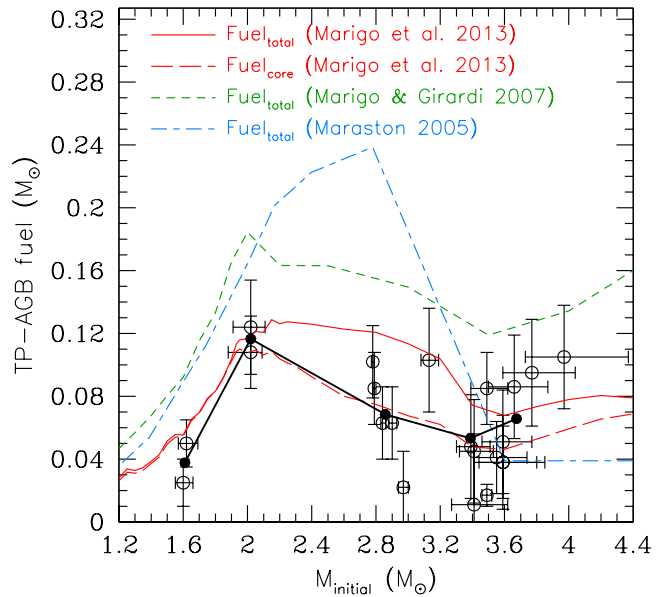}}
\caption{The amount of fuel burnt during the TP-AGB.  The red curves are taken from 
the best-fit (see \S~\ref{sec_3dup}) model of Marigo et~al.\ (2013), 
for $Z_{\rm initial}$ = 0.02 and the calibrated $\lambda_{\rm max}$ function  
for the maximum efficiency of the third dredge-up.  The long-dashed curve 
shows the TP-AGB fuel related to the net core-mass growth (i.e., compare to data 
points in black), whereas the solid curve shows the total TP-AGB fuel (e.g., also 
including the part of the fuel that escapes the star in the form of  chemical yields).  
For the best-fit model, the fuel burnt through the 
core-mass growth alone is 90 -- 65\% of the total TP-AGB fuel at $M_{\rm initial}$ = 2 -- 
3.5~$M_\odot$.  As a comparison, the total fuel burnt on the TP-AGB 
in the \cite{maraston05}  (long-dashed short-dashed blue curve) and 
\citet{marigogirardi_07} (short-dashed green line) models are also shown.
Both predictions are, to different extents, significantly larger than 
our best-fit model would indicate.  This fuel is directly proportional to the energy 
output during the TP-AGB phase, which we illustrate in Figure~\ref{fig:lum}. 
\label{fig:fuel}}
\label{fig_fuelcal}
\end{figure}


As discussed above, the new TP-AGB evolutionary models in \cite{marigo13} present several 
advances over previous generation models \citep[e.g.,][]{marigogirardi_07}, and are found 
to be in excellent agreement with the independent observations in the present study.  In 
the discussion that follows, we reference core-mass growth and associated yields based 
on this best-fitting model from Marigo et~al.\ (2013), with calibrated $\lambda_{\rm max}$ 
relation, shown in Figures~\ref{fig_lmax} and \ref{fig_3dup}.  The corresponding evolutionary 
lifetime of TP-AGB stars with $Z_{\rm initial}$ = 0.02 are illustrated in Figure~\ref{fig:lifetime}.  
The lifetime of stars on the TP-AGB increases rapidly from $\tau$ = 1.4 to 3.4~Myr for stars 
with $M_{\rm initial}$ = 1.6 to $\sim$1.95~$M_\odot$, and then decreases to $\tau$ $\sim$ 2~Myr for 
$M_{\rm initial}$ = 2.5~$M_\odot$, $\tau$ $\sim$ 1~Myr for $M_{\rm initial}$ = 3.0~$M_\odot$, and 
$\tau$ $\sim$ 0.45~Myr for $M_{\rm initial}$ = 3.5~$M_\odot$.

\begin{figure}[h]
\resizebox{1.0\hsize}{!}{\includegraphics{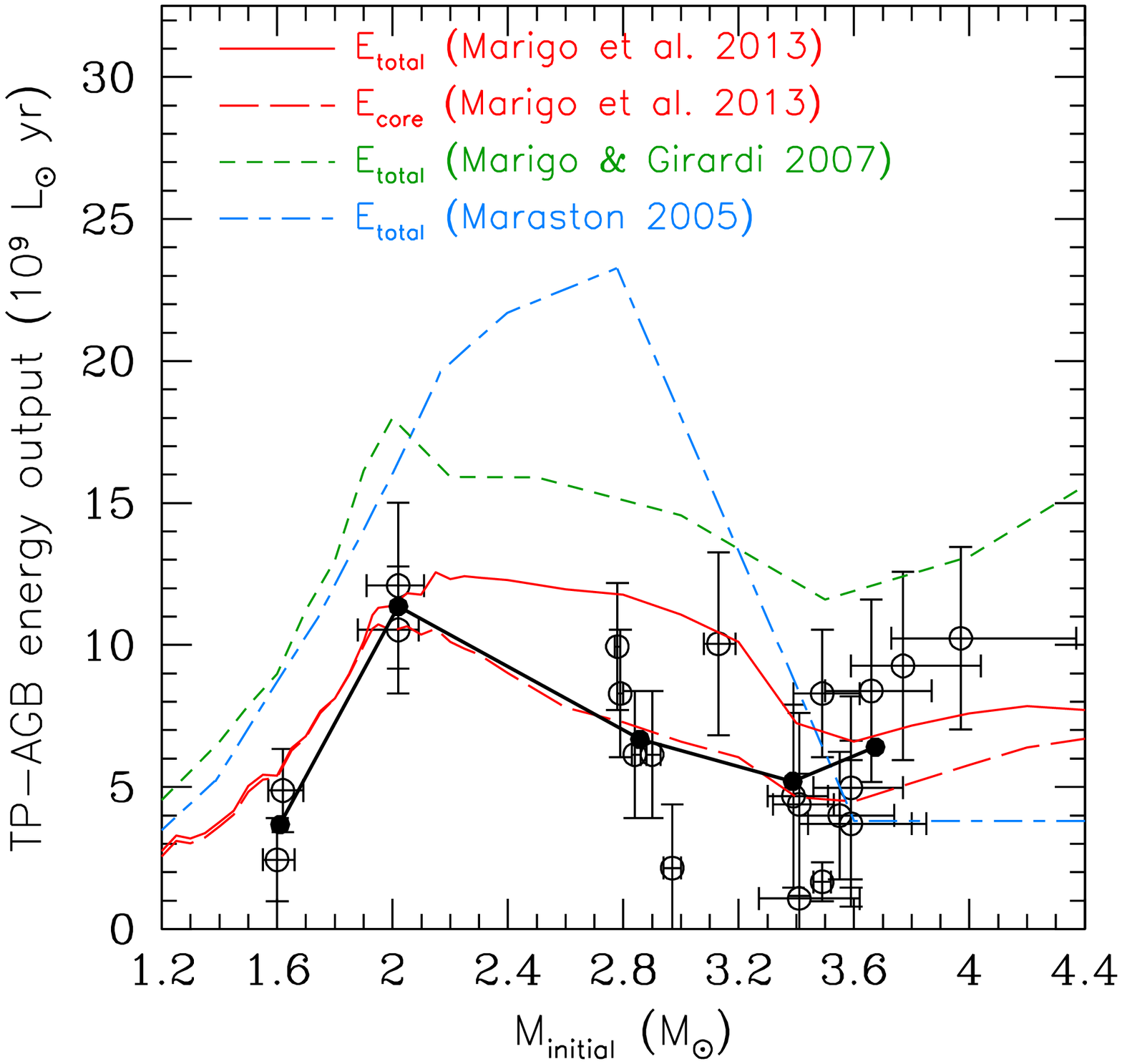}}
\caption{The derived TP-AGB energy output from the best-fit model discussed earlier, both for 
the energy that results from the net core mass growth (dashed red curve) and the total 
energy (solid red curve).  The black data points and black solid curve illustrate the 
new observational constraints from our study, which agree nicely with this model.  The TP-AGB 
energy output is $E$ = 12 $\times$ 10$^{9}$~$L_\odot$~yr for stars with $M_{\rm initial}$ 
$\sim$ 2~$M_\odot$, and steadily decreases for higher mass stars down to $E$ = 6.1 $\times$ 
10$^{9}$~$L_\odot$~yr for stars with $M_{\rm initial}$ $\sim$ 3.5~$M_\odot$. \label{fig:lum}}
\label{fig_enercal}
\end{figure}


\begin{table*}
\begin{center}
\caption{Best-Fitting TP-AGB Model from Marigo et~al.\ (2013)}
\begin{tabular}{cccccccccc}
\hline
\hline
\multicolumn{1}{c}{$M_{\rm initial}$} &
\multicolumn{1}{c}{$M_{\rm{ c, 1tp}}$} &
\multicolumn{1}{c}{$M_{\rm final}$} &
\multicolumn{1}{c}{Fuel$_{\rm core}$ (Fuel$_{\rm total}$)} &
\multicolumn{1}{c}{$t_{\rm TP-AGB}$} &
\multicolumn{1}{c}{$E_{\rm core}$ ($E_{\rm total}$)} &
\multicolumn{1}{c}{Fuel$^*_{\rm total}$} &
\multicolumn{1}{c}{$t^*_{\rm TP-AGB}$} &
\multicolumn{1}{c}{$E^*_{\rm total}$} &
\multicolumn{1}{c}{C/O$_{\rm final}$} \\
\multicolumn{1}{c}{($M_\odot$)} &
\multicolumn{1}{c}{($M_\odot$)} &
\multicolumn{1}{c}{($M_\odot$)}  &
\multicolumn{1}{c}{($M_\odot$)} &
\multicolumn{1}{c}{(Myr)} &
\multicolumn{1}{c}{(10$^{9}$$L_\odot$ yr)} &
\multicolumn{1}{c}{($M_\odot$)} &
\multicolumn{1}{c}{(Myr)} &
\multicolumn{1}{c}{(10$^{9}$$L_\odot$ yr)} &
\multicolumn{1}{c}{photosphere} \\
\hline
    1.00 &    0.518 &    0.541 &    0.018 (0.018) &    0.785 &    1.77 (1.77) & 0.015 &    0.456 & 1.44 &  0.474 \\
    1.05 &    0.521 &    0.543 &    0.017 (0.017) &    0.679 &    1.66 (1.66) &  0.014 &    0.424 & 1.41 & 0.463 \\
    1.10 &    0.523 &    0.546 &    0.018 (0.018) &    0.678 &    1.78 (1.78) &  0.016 &    0.442 & 1.56 & 0.452 \\
    1.15 &    0.522 &    0.555 &    0.026 (0.026) &    0.989 &    2.54 (2.54) &  0.022 &    0.618 & 2.17 & 0.442 \\
    1.20 &    0.524 &    0.558 &    0.026 (0.026) &    0.911 &    2.56 (2.56) &  0.023 &    0.606 & 2.26 & 0.431 \\
    1.25 &    0.524 &    0.565 &    0.032 (0.032) &    1.099 &    3.11 (3.11) &  0.028 &    0.724 & 2.74 & 0.423 \\
    1.30 &    0.526 &    0.566 &    0.031 (0.031) &    0.980 &    3.03 (3.03) &  0.028 &    0.693 & 2.77 & 0.416 \\
    1.35 &    0.528 &    0.570 &    0.033 (0.033) &    0.978 &    3.23 (3.23) &  0.031 &    0.716 & 3.00 & 0.410 \\
    1.40 &    0.529 &    0.576 &    0.037 (0.037) &    1.069 &    3.64 (3.64) &  0.035 &    0.794 & 3.39 & 0.403 \\
    1.45 &    0.529 &    0.582 &    0.041 (0.041) &    1.156 &    4.04 (4.04) &  0.039 &    0.869 & 3.78 & 0.395 \\
    1.50 &    0.524 &    0.588 &    0.050 (0.050) &    1.471 &    4.86 (4.86) &  0.045 &    1.058 & 4.45 & 0.388 \\
    1.55 &    0.524 &    0.594 &    0.054 (0.054) &    1.526 &    5.28 (5.28) &  0.050 &    1.126 & 4.88 & 0.380 \\
    1.60 &    0.528 &    0.598 &    0.055 (0.055) &    1.390 &    5.35 (5.35) &  0.052 &    1.096 & 5.08 & 0.379 \\
    1.65 &    0.521 &    0.604 &    0.064 (0.064) &    1.734 &    6.31 (6.31) &  0.060 &    1.304 & 5.89 & 0.383 \\
    1.70 &    0.520 &    0.609 &    0.069 (0.069) &    1.744 &    6.76 (6.76) &  0.065 &    1.368 & 6.41 & 0.383 \\
    1.75 &    0.514 &    0.615 &    0.078 (0.078) &    2.069 &    7.64 (7.64) &  0.073 &    1.545 & 7.12 & 0.383 \\
    1.80 &    0.513 &    0.621 &    0.083 (0.083) &    2.081 &    8.16 (8.16) &  0.079 &    1.621 & 7.73 & 0.383 \\
    1.85 &    0.507 &    0.627 &    0.092 (0.092) &    2.336 &    8.98 (8.98) &  0.086 &    1.765 & 8.43 & 0.384 \\
    1.90 &    0.499 &    0.632 &    0.102 (0.102) &    2.896 &   10.01 (10.01) &  0.091 &    1.847 & 8.87 & 0.385 \\
    1.93 &    0.493 &    0.634 &    0.108 (0.111) &    3.314 &   10.62 (10.87) &  0.095 &    1.944 & 9.35 & 0.494 \\
    1.95 &    0.491 &    0.634 &    0.110 (0.114) &    3.426 &   10.77 (11.14) &  0.097 &    1.968 & 9.52 & 0.541 \\
    2.00 &    0.498 &    0.638 &    0.108 (0.115) &    3.148 &   10.55 (11.30) &  0.103 &    2.014 & 10.05 & 0.673 \\
    2.05 &    0.499 &    0.642 &    0.109 (0.121) &    3.125 &   10.71 (11.84) &  0.109 &    2.089 & 10.71 & 0.789 \\
    2.10 &    0.508 &    0.647 &    0.106 (0.122) &    2.842 &   10.42 (11.90) &  0.113 &    2.080 & 11.08 & 0.884 \\
    2.15 &    0.511 &    0.653 &    0.108 (0.129) &    2.825 &   10.62 (12.68) &  0.123 &    2.183 & 12.00 & 1.034 \\
    2.20 &    0.515 &    0.651 &    0.104 (0.128) &    2.712 &   10.16 (12.57) &  0.123 &    2.162 & 12.00 & 1.101 \\
    2.25 &    0.517 &    0.650 &    0.101 (0.129) &    2.666 &    9.91 (12.66) &  0.124 &    2.182 & 12.17 & 1.155 \\
    2.30 &    0.524 &    0.654 &    0.099 (0.129) &    2.467 &    9.70 (12.66) &  0.126 &    2.113 & 12.32 & 1.185 \\
    2.40 &    0.534 &    0.655 &    0.092 (0.129) &    2.298 &    9.05 (12.60) &  0.126 &    2.048 & 12.38 & 1.258 \\
    2.60 &    0.564 &    0.670 &    0.080 (0.126) &    1.815 &    7.84 (12.31) &  0.125 &    1.758 & 12.26 & 1.321 \\
    2.80 &    0.597 &    0.697 &    0.075 (0.124) &    1.465 &    7.32 (12.15) &  0.124 &    1.465 & 12.15 & 1.290 \\
    3.00 &    0.637 &    0.726 &    0.068 (0.117) &    1.125 &    6.62 (11.43) &  0.117 &    1.125 & 11.43 & 1.208 \\
    3.20 &    0.681 &    0.763 &    0.062 (0.104) &    0.822 &    6.08 (10.19) &  0.104 &    0.822 & 10.19 & 1.075 \\
    3.40 &    0.724 &    0.787 &    0.048 (0.074) &    0.509 &    4.67 (7.28) &  0.074 &    0.509 & 7.28 & 0.797 \\
    3.60 &    0.751 &    0.812 &    0.046 (0.067) &    0.405 &    4.52 (6.58) &  0.067 &    0.405 & 6.58 & 0.710 \\
    3.80 &    0.762 &    0.832 &    0.053 (0.073) &    0.409 &    5.16 (7.16) &  0.073 &    0.409 & 7.16 & 0.711 \\
    4.00 &    0.773 &    0.853 &    0.059 (0.078) &    0.404 &    5.80 (7.60) &  0.078 &    0.404 & 7.60 & 0.686 \\
    4.20 &    0.786 &    0.875 &    0.066 (0.081) &    0.388 &    6.42 (7.90) &  0.081 &    0.388 & 7.90 & 0.643 \\
    4.40 &    0.803 &    0.898 &    0.069 (0.079) &    0.353 &    6.73 (7.78) &  0.079 &    0.353 & 7.78 & 0.573 \\
\hline
\multicolumn{8}{l}{* Quantities integrated for luminosities $\log (L/L_{\odot}) > 3.4$, i.e. brighter than the RGB tip.}\\
\end{tabular}
\label{table3}
\end{center}
\end{table*}


\begin{table*}
\begin{center}
\caption{Measurements of TP-AGB Core-Mass Growth, Fuel, and Energy Output}
\begin{tabular}{ccccccc}
\hline
\hline
\multicolumn{1}{c}{$M_{\rm initial}$} & \multicolumn{1}{c}{$M_{\rm final}$} &
\multicolumn{1}{c}{$M_{\rm{ c, 1tp}}$} & \multicolumn{1}{c}{$\Delta M_{\rm growth}$} & 
\multicolumn{1}{c}{Fuel$_{\rm core}$} & \multicolumn{1}{c}{$E_{\rm core}$} \\
\multicolumn{1}{c}{($M_\odot$)} & \multicolumn{1}{c}{($M_\odot$)} & \multicolumn{1}{c}{($M_\odot$)}  & 
\multicolumn{1}{c}{($M_\odot$)} & \multicolumn{1}{c}{($M_\odot$)} & \multicolumn{1}{c}{(10$^{9}$$L_\odot$ yr)} \\
\hline
1.60$^{+0.06}_{-0.05}$ & 0.560  & 0.528  & 0.032 $\pm$ 0.020  & 0.025 $\pm$ 0.015  &  2.43 $\pm$ 1.52 \\ 
1.62$^{+0.07}_{-0.05}$ & 0.590  & 0.525  & 0.065 $\pm$ 0.020  & 0.050 $\pm$ 0.015  &  4.93 $\pm$ 1.52 \\ 
2.02$^{+0.07}_{-0.14}$ & 0.640  & 0.498  & 0.142 $\pm$ 0.030  & 0.108 $\pm$ 0.023  & 10.61 $\pm$ 2.24 \\ 
2.02$^{+0.09}_{-0.11}$ & 0.660  & 0.498  & 0.162 $\pm$ 0.040  & 0.124 $\pm$ 0.030  & 12.11 $\pm$ 2.99 \\ 
2.78$^{+0.01}_{-0.01}$ & 0.730  & 0.594  & 0.136 $\pm$ 0.030  & 0.102 $\pm$ 0.023  & 10.04 $\pm$ 2.21 \\ 
2.79$^{+0.01}_{-0.01}$ & 0.710  & 0.597  & 0.113 $\pm$ 0.030  & 0.085 $\pm$ 0.023  &  8.34 $\pm$ 2.21 \\ 
2.84$^{+0.02}_{-0.01}$ & 0.690  & 0.606  & 0.084 $\pm$ 0.030  & 0.063 $\pm$ 0.023  &  6.20 $\pm$ 2.21 \\ 
2.90$^{+0.03}_{-0.02}$ & 0.700  & 0.617  & 0.083 $\pm$ 0.030  & 0.063 $\pm$ 0.023  &  6.13 $\pm$ 2.21 \\ 
2.97$^{+0.03}_{-0.03}$ & 0.660  & 0.631  & 0.029 $\pm$ 0.030  & 0.022 $\pm$ 0.023  &  2.14 $\pm$ 2.22 \\ 
3.13$^{+0.06}_{-0.05}$ & 0.802  & 0.665  & 0.137 $\pm$ 0.043  & 0.103 $\pm$ 0.033  & 10.13 $\pm$ 3.18 \\ 
3.39$^{+0.12}_{-0.09}$ & 0.785  & 0.721  & 0.064 $\pm$ 0.043  & 0.048 $\pm$ 0.033  &  4.74 $\pm$ 3.19 \\ 
3.41$^{+0.12}_{-0.09}$ & 0.785  & 0.726  & 0.059 $\pm$ 0.043  & 0.045 $\pm$ 0.033  &  4.37 $\pm$ 3.19 \\ 
3.41$^{+0.21}_{-0.14}$ & 0.740  & 0.726  & 0.014 $\pm$ 0.060  & 0.011 $\pm$ 0.045  &  1.04 $\pm$ 4.44 \\ 
3.49$^{+0.13}_{-0.10}$ & 0.850  & 0.737  & 0.113 $\pm$ 0.030  & 0.085 $\pm$ 0.023  &  8.35 $\pm$ 2.22 \\ 
3.49$^{+0.03}_{-0.03}$ & 0.760  & 0.737  & 0.023 $\pm$ 0.010  & 0.017 $\pm$ 0.007  &  1.70 $\pm$ 0.74 \\ 
3.55$^{+0.19}_{-0.14}$ & 0.800  & 0.745  & 0.055 $\pm$ 0.030  & 0.041 $\pm$ 0.023  &  4.06 $\pm$ 2.21 \\ 
3.59$^{+0.18}_{-0.13}$ & 0.817  & 0.749  & 0.068 $\pm$ 0.044  & 0.051 $\pm$ 0.033  &  5.01 $\pm$ 3.24 \\ 
3.59$^{+0.26}_{-0.18}$ & 0.800  & 0.749  & 0.051 $\pm$ 0.040  & 0.038 $\pm$ 0.030  &  3.76 $\pm$ 2.95 \\ 
3.59$^{+0.21}_{-0.15}$ & 0.800  & 0.749  & 0.051 $\pm$ 0.030  & 0.038 $\pm$ 0.023  &  3.76 $\pm$ 2.21 \\ 
3.66$^{+0.21}_{-0.16}$ & 0.869  & 0.754  & 0.115 $\pm$ 0.044  & 0.086 $\pm$ 0.033  &  8.44 $\pm$ 3.23 \\ 
3.77$^{+0.27}_{-0.18}$ & 0.888  & 0.760  & 0.128 $\pm$ 0.045  & 0.095 $\pm$ 0.034  &  9.33 $\pm$ 3.28 \\ 
3.97$^{+0.40}_{-0.24}$ & 0.914  & 0.771  & 0.143 $\pm$ 0.045  & 0.105 $\pm$ 0.033  & 10.30 $\pm$ 3.24 \\ 
\hline
\end{tabular}
\label{table4}
\end{center}
\end{table*}



The peak in the TP-AGB lifetime takes place in correspondence to the stellar progenitor 
whose mass is the closest to the maximum mass, $M_{\rm HeF}$, for a star to experience 
the He-flash in the degenerate core at the tip of the RGB.  In fact, for $M_{\rm initial} 
\simeq M_{\rm HeF}$ stellar evolution models expect a minimum in the core mass at the 
first thermal pulse \citep[e.g.,][]{lattanzio_86, bressan12}.  Therefore, stars with 
initial masses close to this limit enter the TP-AGB phase at fainter luminosities
 compared to their neighbors in mass, normally below the tip of the RGB.
The net effect is a longer duration of the TP-AGB phase just in proximity of 
$M_{\rm HeF}$, that is $\simeq 1.95\,M_{\odot}$ for the chemical composition considered 
here.

The energy output provided by a star during its TP-AGB phase is simply the time integral 
of the luminosity over the TP-AGB lifetime, and is proportional to the total amount of 
nuclear fuel burnt during the evolutionary phase \citep{renzini86}.  More recent studies 
\citep{marigogirardi_01, bird11} have pointed out that the core-mass growth on the TP-AGB 
provides only a {\em lower limit} to the total fuel consumption, since part of the nuclear 
fuel may either be taken away from the core by dredge-up events, or occur outside the core, 
like in the case of hot-bottom burning in more massive AGB stars.  The part of nuclear 
fuel not locked in the core is eventually lost by the stars in the form of chemical yields, 
as extensively discussed in \cite{marigogirardi_01}.

In Figure~\ref{fig_fuelcal}, we illustrate the fuel burnt on the TP-AGB from the best-fit 
model, both for the fuel just related to the growth of the stellar core (dashed red curve) 
and the total fuel (solid red curve).  From near the peak fuel consumption at 
$M_{\rm initial}$ $\sim$ 2~$M_\odot$ to 3.5~$M_\odot$, the core-mass growth accounts 
for 90 to 65\% of the total TP-AGB fuel.  The model predictions for the amount of 
fuel burnt through the core-mass growth are in excellent agreement with our data points 
(black points and  solid line).  For this set of calculations  we find that the fraction 
of the total fuel expelled in the form of chemical yields is zero for $M_{\rm initial} 
\lesssim 1.9~$M$_\odot$, then it increases up to $\simeq 40\%$ for $M_{\rm initial}$ $\sim$ 
3~$M_\odot$, and finally decreases to $\simeq 25\%$ for $M_{\rm initial}$ $\sim$ 4~$M_\odot$.

For comparison, we also illustrate the total TP-AGB 
fuel predicted by the \cite{marigogirardi_07} and \cite{maraston05} models, 
for $Z_{\rm initial}=0.019$ and $Z_{\rm initial}=0.02$, respectively. Both curves are 
higher than the total fuel expected from our best-fit set of TP-AGB calculations.
At initial masses $M_{\rm initial}$ $\sim$ 1.6, 2.0, 2.8, 3.0~$M_\odot$ the 
\cite{marigogirardi_07} and \cite{maraston05} models exceed our calibrated TP-AGB fuel 
roughly by 65\%,  57\%, 26\%, 66\%, and 61\%,  41\%, 90\%, 16\%,  respectively.

Following the prescription in \cite{marigogirardi_01}, it is straightforward to convert the 
amount of fuel burnt through the core-mass growth to establish a lower limit of the 
integrated luminosity emitted during the TP-AGB phase.  This result depends only on the 
measured core-mass growth, the efficiency of H-burning reactions ($A_{\rm H}$ = 9.79 $\times$ 
10$^{10}$~$L_\odot$~yr, Marigo \& Girardi 2001), and the surface abundance of H.  The results 
are illustrated in Figure~\ref{fig:lum}.  As above, the red dashed curve is the output 
energy associated with just the core-mass growth and is in excellent agreement with the 
data (black points and black solid curve).   The solid red curve is the same model, but for 
the total energy.  The TP-AGB energy output is  therefore 
$E \simeq 11-12 \times 10^{9}$~$L_\odot$~yr for stars with $2\,M_{\odot} \la M_{\rm initial} 
\la 3\, M_{\odot}$, and then decreases for higher mass stars down to $E$ = 6 -- 7  
$\times$ 10$^{9}$~$L_\odot$~yr for stars with $3.5\,M_{\odot} \la M_{\rm initial} \la 4.5\, M_{\odot}$.

We present theoretical predictions of the TP-AGB core mass at the first thermal pulse, final 
mass at the end of the TP-AGB, fuel consumed, stellar lifetime, and stellar energy output, 
final surface C/O ratio, based on the best-fitting model from  Marigo et~al.\ (2013) in 
Table~\ref{table3}.  In Table~4, we derive these quantities, other than C/O, for each of the 
stars in our data set.


%



\section{Conclusion} \label{conclusion}

The physical processes occurring on the TP-AGB phase of stellar evolution lead to dynamic changes 
in the nature of stars.  Over the course of just a few million years, stars can shed $>$75\% of 
their mass through winds during this evolution.  The theoretical parametrization of these 
processes plays a critical role in the interpretation of light from unresolved galaxies 
(especially at intermediate ages), however, such efforts are relatively unconstrained by 
observations.  In this paper, we leverage new discoveries of white dwarfs in the nearby and 
well-studied Hyades and Praesepe star clusters to establish 18 initial and final mass pairs, 
combined with earlier studies by our team of the older star clusters NGC~6819 and NGC~7789.  
These data provide new insights on the properties of the TP-AGB phase of stellar evolution.  

We measure the growth of the core mass on the TP-AGB to be 10\% at $M_{\rm initial}$ = 1.6, 
rising rapidly to 30\% at $M_{\rm initial} \simeq 2.0\, M_{\odot}$.  For more massive stars, 
the core-mass growth is lower and decreases steadily to $\sim$10\% at $M_{\rm initial} \simeq 
3.4\, M_{\odot}$.  These results are in nice agreement with the new TP-AGB models in 
\cite{marigo13} for initial metallicity $Z_{\rm initial}=0.02$, 
which offer several advances over previous generation calculations.  
By comparing to models with varying efficiencies of the third dredge-up and 
different mass-loss prescriptions, we demonstrate 
that the stellar mass loss rate plays the dominant role in guiding the core-mass growth, but 
the third dredge-up also produces an important effect that must be taken into account.

We find that the semi-empirical \citet{Bedijn_88}-like mass-loss relation (adopted in 
Marigo et~al.\ 2013) and the \citet{VassiliadisWood_93} formula yield 
a very good agreement with the new white dwarfs mass measurements, while other 
prescriptions in the literature need to be tuned by adjusting ad-hoc multiplicative factors.  
Our exploratory calibration (see Figure~\ref{fig_lmax}) suggests to adopt  
$\eta_{\rm B}\simeq 0.05$ in the \citet{Blocker95} formula, $\eta_{\rm vL}\simeq 0.4$ 
in the \citet{vanLoon_etal05}  relation, and  $\eta_{\rm R}\simeq 2$ in 
the \citet{Reimers75} law.  We note, however, that this latter law produces less 
satisfactory results, failing to reproduce the morphology of the observed relation
between the core-mass growth and the initial stellar mass, and in general, it should
not be considered a suitable choice for the TP-AGB phase.

A tentative calibration of the third dredge-up efficiency at metallicities  
$Z_{\rm initial}=0.02$, as a function of the stellar mass  would indicate
that i) stars with  $M_{\rm initial}$ $<$ 1.9~$M_{\odot}$ do not experience 
the third dredge-up,  in agreement with predictions of full AGB models 
\citep{karakas02}; ii) at larger masses the efficiency of the third dredge-up 
increases quickly with the stellar mass up to values $\lambda_{\rm max}$ $\simeq$ 0.5 
for $M_{\rm initial}$ $\simeq$ 2.5 -- 3.0~$M_{\odot}$; iii) this positive   
trend is eventually reversed and the third dredge-up becomes less efficient 
with increasing stellar mass, illustrating a larger core-mass growth.  The latter point is at odds with full TP-AGB 
models \citep{karakas02} that predict $\lambda_{\rm max}$ $\simeq$ 0.9 -  1.0 for 
$M_{\rm initial}$ $<$ 4.0~$M_{\odot}$.  Given its critical impact on the chemical 
yields from more massive AGB stars, this aspect demands a further careful 
analysis, which is postponed to a follow-up work.  In any case, the inefficient 
C-star formation at $Z_{\rm initial}$ = 0.02, that follows from this preliminary 
calibration, is supported by the recent study of \citet{Boyer_etal13}, who have 
pointed out a dramatic scarcity of C stars in the inner disk of the M31 galaxy, 
a region characterized by a metallicity comparable to that considered in this 
work ([Fe/H] $\simeq$ $+$0.1).
 
Finally, we relate the core-mass growth to the nuclear fuel burnt during the 
TP-AGB phase to calculate the energy output of stars in this phase as summarized 
in Tables~\ref{table3} (best-fitting model) and \ref{table4} (data).  
At the peak core-mass growth for stars with $M_{\rm initial}$ $\sim$ 2~$M_\odot$, 
the TP-AGB lifetime is $\tau$ $\simeq$ 3.4~Myr, which reduces to $\tau$ $\simeq$ 2~Myr for 
luminosities brighter than the RGB tip (i.e., $\log(L/L_{\odot})$ $>$ 3.4). The 
corresponding integrated luminosity is $L$ $\simeq$ 12 $\times$ 10$^{9}$~$L_\odot$~yr.  

Our measurements illustrate that the fuel burnt during the TP-AGB for 
metallicity $Z_{\rm i}$ $\simeq$ 0.02, is substantially lower than adopted by 
\cite{maraston05}, and to a lesser extent, than predicted by \cite{marigogirardi_07}.  
This finding is in line with other recent studies that, from
independent arguments, favor a {\em lighter} TP-AGB contribution to the integrated
galaxy light, \citep[e.g.,][]{Kriek_etal10, melbourne12, zibetti_etal13, conroy13}.  Our 
results are also in line with the recent conclusions of \cite{girardi13}, who point 
out at an insidious problem in present derivations of the TP-AGB fuel based on Magellanic 
Cloud star clusters.

We caution that the conclusions drawn from this study apply to the TP-AGB stars 
with slightly super-solar metallicity, and a straightforward extrapolation 
to lower metallicities is not correct and should be avoided. Accomplishing a thorough 
and reliable TP-AGB calibration requires an observational sampling over the entire 
relevant ranges of ages and metallicites.  Accurate white dwarf mass measurements in 
additional intermediate-aged star clusters, like those presented in this work, provide 
us with a valuable contribution to achieve this ambitious and challenging goal.






\acknowledgements

We wish to thank L.\ Girardi for providing us with the latest stellar evolution models to 
translate stellar lifetimes into masses.  We also wish to thank Jonathan Bird and Marc 
Pinsonneault for several useful discussions, and Brad Hansen for leading the Keck spectroscopic 
proposal that led to the white dwarf measurements in NGC~6819 and NGC~7789.  We acknowledge the 
help of the team in the Kalirai et~al.\ (2008) study for the initial measurements.  This project 
was supported by the National Science Foundation (NSF) through grant AST-1211719.  
P.M.\ acknowledges financial support from {\em Progetto di Ateneo 2012}, 
ID-CPDA125588/12,  University of Padova. P-E.T.\ was supported during this 
project by the Alexander von Humboldt Foundation and by NASA through Hubble Fellowship grant 
HF-51329.01, awarded by the Space Telescope Science Institute, which is operated by the 
Association of Universities for Research in Astronomy, Incorporated, under NASA contract 
NAS5-26555.



\begin{thebibliography}{}
\bibitem[An et~al.(2008)]{an08} An, D., et~al.\ 2008, ApJS, 179, 326

\bibitem[Anthony-Twarog(1982)]{anthony-twarog82} Anthony-Twarog, B.~J.\ 1982, \apj, 255, 245

\bibitem[Anthony-Twarog(1984)]{anthony-twarog84} Anthony-Twarog, B.~J.\ 1982, \aj, 89, 267

\bibitem[Bedijn(1988)]{Bedijn_88} Bedijn, P.~J.\ 1988, \aap, 205, 105 

\bibitem[Bergeron, Saffer, \& Liebert(1992)]{bergeron92} Bergeron, P., 
Saffer, R.~A., \& Liebert, J.\ 1992, \apj, 394, 228

\bibitem[Bird \& Pinsonneault(2011)]{bird11} Bird, J.~C., \& Pinsonneault, M.~H.\ 
2011, \apj, 733, 81

\bibitem[Bl\"ocker(1995)]{Blocker95} Bl\"ocker, T.\ 1995, \aap, 297, 727 

\bibitem[Boyer et al.(2013)]{Boyer_etal13} Boyer, M.~L., Girardi, L., Marigo, P., et al.\ 2013, 
\apj, 774, 83

\bibitem[Bressan et~al.(2012)]{bressan12} Bressan, A., Marigo, P., Girardi, L., Salasnich, B., 
Dal~Cero, C., Rubele, S., \& Nanni A.\ 2012, \mnras in press, arXiv:1208.4498

\bibitem[Bruzual \& Charlot(2003)]{bruzual03} Bruzual, G.~A., \& Charlot, S.\ 2003, \mnras, 344, 1000

\bibitem[Bruzual A.(2010)]{Bruzual_10} Bruzual A., G.\ 2010, IAU Symposium, 262, 55 

\bibitem[Casewell et~al.(2009)]{casewell09} Casewell, S.~L., Dobbie, P.~D., Napiwotzki, R., Burleigh, M.~R., 
Barstow, M.~A., \& Jameson, R.~F.\ 2009, \mnras, 395, 1795

\bibitem[Claver et~al.(2001)]{claver01} Claver, C.~F., Liebert, J., Bergeron, P., 
\& Koester, D.\ 2001, \apj, 563, 987


\bibitem[Conroy \& Gunn(2010)]{conroy10} Conroy, C., \& Gunn, J.~E.\ 2010, \apj, 712, 833

\bibitem[Conroy(2013)]{conroy13} Conroy, C.\ 2013, arXiv:1301.7095 

\bibitem[Cranmer \& Saar(2011)]{Cranmer_11} Cranmer, S.~R., \& Saar, S.~H.\ 2011, \apj, 741, 54 

\bibitem[Dalcanton et~al.(2009)]{dalcanton09} Dalcanton, J., et~al.\ 2009, \apjs, 183, 67

\bibitem[Dobbie et~al.(2004)]{dobbie04} Dobbie, P.~D., Pinfield, D.~J., 
Napiwotzki, R., Hambly, N.~C., Burleigh, M.~R., Barstow, M.~A., 
Jameson, R.~F., \& Hubeny, I.\ 2004, \mnras, 355, L39

\bibitem[Dobbie et~al.(2006)]{dobbie06} Dobbie, P.~D., 
et~al.\ 2006, \mnras, 369, 383

\bibitem[Eggen \& Greenstein(1965)]{eggen65} Eggen, O.~J., \& Greenstein, 
J.~L.\ 1965, \apj, 141, 83

\bibitem[Frogel et al.(1990)]{Frogel_etal90} Frogel, J.~A., Mould, J., \& Blanco, V.~M.\ 1990, \apj, 352, 96 

\bibitem[Girardi \& Marigo(2007)]{girardi07} Girardi, L., \& Marigo, P.\ 2007, 
A\&A, 462, 237

\bibitem[Girardi et~al.(2010)]{girardi10} Girardi~L., et~al.\ 2010, \apj, 
724, 1030

\bibitem[Girardi et~al.(2013)]{girardi13} Girardi, L., Marigo, P., Bressan, A., \& Rosenfield, P.\ 2013, ApJ, 
777, 142

\bibitem[Gratton(2000)]{gratton00} Gratton, R.\ 2000, ASP Conf. Ser., 198, 225

\bibitem[Groenewegen \& de Jong(1993)]{GroenJong_93} Groenewegen, M.~A.~T., \& de Jong, T.\ 1993, \aap, 267, 410 

\bibitem[Groenewegen et al.(2009)]{Groen_etal09} Groenewegen, M.~A.~T., Sloan, G.~C., Soszy{\'n}ski, I., \& Petersen, E.~A.\ 2009, \aap, 506, 1277 

\bibitem[Habing(1996)]{habing96} Habing, H.~J.\ 1996, \araa, 7, 97

\bibitem[Herwig(2005)]{herwig05} Herwig, F.\ 2005, \araa, 43, 435

\bibitem[Herwig(2004)]{herwig04} Herwig, F.\ 2004, ApJ, 605, 425 

\bibitem[Kalirai et~al.(2001)]{kalirai01} Kalirai, J.~S., Richer, H.~B., Fahlman, G., 
Cuillandre, J.-C., Ventura, P., D'Antona, F., Bertin, E., Marconi, G., \& Durrell, P.\ 
2001, \aj, 122, 266

\bibitem[Kalirai et~al.(2005)]{kalirai05} Kalirai, J.~S., Richer, H.~B., Reitzel, 
D., Hansen, B.~M.~S., Rich, R.~M., Fahlman, G.~G., Gibson, B.~K., \& von~Hippel, 
T.\ 2005, \apjl, 618, L123

\bibitem[Kalirai et~al.(2007)]{kalirai07} Kalirai, J.~S., Bergeron, P., 
Hansen, B.~M.~S., Kelson, D.~D., Reitzel, D.~B., Rich, R.~M., \& Richer, 
H.~B.\ 2007, \apj, 671, 748

\bibitem[Kalirai et~al.(2008)]{kalirai08} Kalirai, J.~S., Hansen, B.~M.~S., Kelson, D.~D., 
Reitzel, D.~B., Rich, R.~M., \& Richer, H.~B.\ 2008, \apj, 676, 594

\bibitem[Kalirai et~al.(2009)]{kalirai09} Kalirai, J.~S., Davis, S.~D., Richer, H.~B., 
Bergeron, P., Catelan, M., Hansen, B.~M.~S., \& Rich, R.~M.\ 2009, \apj, 705, 408

\bibitem[Kalirai (2012)]{kalirai12} Kalirai, J.~S.\ 2012, Nature, 486, 90

\bibitem[Kamath et al.(2012)]{Kamath_etal12} Kamath, D., Karakas, A.~I., \& Wood, P.~R.\ 2012, \apj, 746, 20 

\bibitem[Kamath et al.(2010)]{Kamath_etal10} Kamath, D., Wood, P.~R., Soszy{\'n}ski, I., \& Lebzelter, T.\ 2010, \mnras, 408, 522 

\bibitem[Karakas, Lattanzio, \& Pols (2002)]{karakas02} Karakas, A.~I., Lattanzio, J.~C., \& 
Pols, O.\ 2002, Publications of the Astronomical Society of Australia, 19, 515

\bibitem[Kleinman et~al.(2013)]{kleinman13} Kleinman, S.~J., et~al.\ 2013, \apjs, 204, 5

\bibitem[Kriek et al.(2010)]{Kriek_etal10} Kriek, M., Labb{\'e}, I., Conroy, C., et al.\ 2010, \apjl, 722, L64 

\bibitem[Lambert et al.(1986)]{Lambert_etal86} Lambert, D.~L., Gustafsson, B., Eriksson, K., \& Hinkle, K.~H.\ 1986, \apjs, 62, 373 

\bibitem[Lattanzio(1986)]{lattanzio_86} Lattanzio, J.~C.\ 1986, \apj, 311, 708 

\bibitem[Luyten(1962)]{luyten62} Luyten, W.~J.\ 1962, The Observatory., Univ.\ Minnesota, 31, 1

\bibitem[Maraston(2005)]{maraston05} Maraston, C.\ 2005, \mnras, 362, 799

\bibitem[Maraston et~al.(2006)]{maraston06} Maraston, C., Daddi, E., Renzini, A., Cimatti, A., 
Dickinson, M., Papovich, C., Pasquali, A., \& Pirzkal, N.\ 2006, \apj, 652, 85

\bibitem[Marigo et~al.(1999)]{marigo99} Marigo, P., Girardi, L., \& Bressan, A.\ 1999, \aap, 344, 123 

\bibitem[Marigo(2002)]{marigo02} Marigo, P.\ 2002, \aap, 387, 507 

\bibitem[Marigo \& Girardi(2001)]{marigogirardi_01} Marigo, P., \& Girardi, L.\ 2001, 
A\&A, 377, 132

\bibitem[Marigo \& Girardi(2007)]{marigogirardi_07} Marigo P., \& Girardi L.\ 
2007, A\&A, 469, 239

\bibitem[Marigo et~al.(2008)]{marigo08} Marigo P., Girardi, L., Bressan, A., Groenewegen, M.~A.~T., 
Aringer, B., Silva, L., \& Granato, G.~L.\ 2008, \aap, 482, 883

\bibitem[Marigo \& Aringer(2009)]{marigo09} Marigo P., \& Aringer B.\ 
2009, A\&A, 508, 1539

\bibitem[Marigo(2012)]{marigo12} Marigo, P.\ 2012, IAU Symposium, 283, 87 

\bibitem[Marigo et al.(2013)]{marigo13} Marigo, P., Bressan, A., Nanni, A., Girardi, L., \& Pumo, M.~L.\ 2013, 
\mnras, 434, 488

\bibitem[Marigo(2013)]{marigo13b} Marigo, P.\ 2013, IAU Symposium, 281, 36 

\bibitem[Mattsson et al.(2010)]{mattsson_etal10} Mattsson, L., Wahlin, R., {\ H\"o}fner, S.\ 2010, \aap, 509, A14 

\bibitem[Melbourne et~al.(2012)]{melbourne12} Melbourne, J., et~al.\ 2012, \apj, 748, 47

\bibitem[Miglio et al.(2012)]{Miglio_etal12} Miglio, A., Brogaard, K., Stello, D., et al.\ 2012, \mnras, 419, 2077 

\bibitem[Ohnaka et al.(2000)]{Ohnaka_etal00} Ohnaka, K., Tsuji, T., \& Aoki, W.\ 2000, \aap, 353, 528 

\bibitem[Perryman et~al.(1998)]{perryman98} Perryman, M.~A.~C., et~al.\ 1998, 
A\&A, 331, 81

\bibitem[Reid(1992)]{reid92} Reid, N.\ 1992, \mnras, 257, 257 

\bibitem[Reimers(1975)]{Reimers75} Reimers, D.\ 1975, Memoires of the Societe Royale des Sciences de Liege, 8, 369 

\bibitem[Renzini \& Voli(1981)]{RenziniVoli_81} Renzini, A., \& Voli, M.\ 1981, \aap, 94, 175 

\bibitem[Renzini \& Buzzoni(1986)]{renzini86} Renzini, A., \& Buzzoni, A.\ 1986, in Astrophysics 
and Space Science Library, Vol. 122, Spectral Evolution of Galaxies, ed.\ C.\ Chiosi \& A.\ Renzini, 
195-231

\bibitem[Schilbach \& R{\"o}ser(2012)]{schilbach12} Schilbach, E. \& R{\"o}ser, S.\ 2012, \aap, 537, A129

\bibitem[Schr{\"o}der \& Cuntz(2005)]{SchroederCuntz_05} Schr{\"o}der, K.-P., \& Cuntz, M.\ 2005, \apjl, 630, L73 

\bibitem[Tremblay \& Bergeron(2009)]{tremblay09} Tremblay, P.-E., \& Bergeron, P.\ 2009, \apj, 696, 1755

\bibitem[Tremblay et~al.(2012)]{tremblay12} Tremblay, P.-E., Schilbach, E., Roser, S., Jordan, S., 
Ludwig, H.-G., \& Goldman, B.\ 2012, \aap, 547, 99

\bibitem[van~Altena(1969)]{vanaltena69} van Altena, W.~F.\ 1969, \aj, 74, 2

\bibitem[van Loon et al.(2005)]{vanLoon_etal05} van Loon, J.~T., Cioni, M.-R.~L., Zijlstra, A.~A., \& Loup, C.\ 2005, \aap, 438, 273 

\bibitem[Vassiliadis \& Wood(1993)]{VassiliadisWood_93} Vassiliadis, E., \& Wood, P.~R.\ 1993, \apj, 413, 641 

\bibitem[Ventura et al.(2000)]{Ventura_etal00} Ventura, P., D'Antona, F., \& Mazzitelli, I.\ 2000, \aap, 363, 605 

\bibitem[von Hippel(1998)]{vonhippel98} von Hippel, T.\ 1998, \aj, 115, 1536

\bibitem[Weidemann et~al.(1992)]{weidemann92} Weidemann, V., Jordan, S., Iben, 
I., Jr., \& Casertano, S.\ 1992, \aj, 104, 1876 

\bibitem[Weidemann(2000)]{weidemann00} Weidemann, V.\ 2000, \aap, 363, 647

\bibitem[Willson(2000)]{willson00} Willson, L.~A.\ 2000, ARA\&A, 38, 573

\bibitem[Zibetti et al.(2013)]{zibetti_etal13} Zibetti, S., Gallazzi, A., Charlot, S., Pierini, D., \& Pasquali, A.\ 2013, \mnras, 428, 1479 

\bibitem[Zuckerman et~al.(2013)]{zuckerman13} Zuckerman, B., Klein, B., Xu, S., \& 
Jura, M.\ 2013, ApJ in press, arXiv:1304.2638

\end{thebibliography}
\end{document}